\documentclass[
    reprint,
    amsmath,
    amssymb,
    aps,
    pra,
    floatfix,
    longbibliography
]{revtex4-2}

\usepackage{amssymb}
\usepackage{graphicx}
\usepackage{tikz}
\usetikzlibrary{calc}
\usepackage{pgfplots}
\usetikzlibrary{arrows.meta}

\pgfplotsset{compat=1.18}
\usepackage{dcolumn}
\usepackage{bm}
\usepackage[T1]{fontenc}
\usepackage[utf8]{inputenc}
\usepackage[final]{microtype}

\usepackage{mathtools}
\usepackage{amsthm}
\usepackage{braket}
\usepackage{siunitx}
\usepackage{csquotes}
\usepackage{mathrsfs}
\usetikzlibrary{fadings}

\newtheorem{theorem}{Theorem}
\newtheorem{proposition}{Proposition}

\newtheorem{definition}{Definition}

\usepackage[section]{placeins}

\usepackage{hyperref}
\usepackage[nameinlink,noabbrev]{cleveref}

\makeatletter
\def\@fnsymbol#1{\ifcase#1\or 1\or 2\or 3\or 4\or 5\or 6\or 7\or 8\or 9\or 10\else\@ctrerr\fi}
\makeatother
\def\alg{\mathsf}
\def\QCA{\textbf}
\newcommand{\proj}[1]{|#1\rangle\!\langle#1|}

\begin{document}

\title{Local Measurement Theory for Quantum Cellular Automata}

\author{Tommaso Brambilla}
\email{tommaso.brambilla02@universitadipavia.it}
\author{Alessandro Tosini}
\email{alessandro.tosini@unipv.it}
\author{Paolo Perinotti}
\email{paolo.perinotti@unipv.it}

\affiliation{Dipartimento di Fisica, Università di Pavia, via Bassi 6, 27100 Pavia, Italy}

\begin{abstract}
The formulation of quantum measurements compatible with relativistic causality is obstructed by the tension between the locality of observable algebras and the dynamical state transformations induced by measurement processes. We develop a local measurement theory for quantum cellular automata, lattices of finite-dimensional quantum systems with a causal reversible dynamics evolving in discrete time steps. Measurement processes are described by localized completely positive maps and organized into operational measurement protocols, finite non-adaptive sequences of local instruments distributed across the causal cone induced by the automaton local dynamics. Each protocol induces an effective measurement on the initial time slice, whose support is constrained by the causal structure generated by the automaton. Non-local measurement effects therefore arise through causal propagation and composition of local operations, rather than from primitive non-local interventions or instantaneous collapse prescriptions. We establish the corresponding spatial support bounds and characterize the resulting measurement architecture. As an explicit example, we construct a two-step protocol for a one-dimensional nearest-neighbour qubit automaton which, for suitable automaton parameters, realizes a Bell-basis projective measurement on two neighbouring sites after coarse-graining of the local outcomes. Away from this point, the same protocol generates a continuous family of deformed Bell measurements whose entanglement and stabilizer Rényi non-classicality can be quantified explicitly. The framework thus provides a causally consistent operator-algebraic mechanism for synthesizing spatially extended quantum measurements from strictly local operations.
\end{abstract}
\maketitle

\section{Introduction}

The problem of formulating a conceptually and operationally coherent \emph{measurement theory} in relativistic quantum mechanics faces problems that were raised in the early 1980's in Ref.~\cite{AharonovAlbert1981} and further studied and analysed in subsequent decades~\cite{Sorkin1993,PhysRevA.49.4331,FewsterVerch2020,BostelmannFewsterRuep2021,BorstenJubbKells2021}. The difficulty is not merely to assign probabilities to observables localized in specified spacetime regions. A measurement is itself a physical process: besides producing an outcome, it induces a transformation of the system, and this transformation must be compatible with the causal structure of the theory. While the problem was raised in a field theoretical context, subsequent analysis separated technical and conceptual issues, showing that abstract quantum measurement theory needs to be enriched in some way when one wants to embody it in a geometric context~\cite{Gisin2024towardsmeasurement}---typically with a resource theory accounting for the implementation cost of a measurement.

In the above perspective, one might want to study the achievability of quantum measurements in the simplest geometric scenario. A recent approach to the foundations of quantum field theories based on discrete spacetime structures~\cite{PhysRevA.90.062106,cite-key} provides then the ideal playground to attempt at a reconciliation of abstract quantum measurement theory with implementation of quantum instruments in a realistic dynamically and geometrically constrained setting. In this scenario, the microscopic substratum of quantum fields is given by Quantum Cellular Automata (QCAs) \cite{SchumacherWerner2004}, that are the quantum counterpart of cellular automata, \emph{i.e.}, arrays of cells provided with a discrete evolution rule that is local---the updated content of a cell is influenced only by its neighbours in one step---and translation-invariant.

In this work, we develop an operator-algebraic framework for the
implementation of quantum measurements in discrete spacetime
systems governed by QCAs. We introduce
Operational Measurement Protocols (OMPs), consisting
of finite sequences of strictly local quantum instruments
interleaved with QCA evolution, and characterize the effective
measurements they induce on a common reference time slice.
We establish that the non-selective dynamics of these
protocols preserves the causal structure of the automaton,
and derive bounds on the spatial support of the resulting
POVMs in terms of the protocol duration and the QCA
neighbourhood. Within this framework, we explicitly
construct a protocol implementing a non-local Bell-basis
projective measurement using only single-site interventions
and nearest-neighbour QCA dynamics. More generally, the
same construction yields a continuous family of deformed
Bell measurements, whose entanglement and stabilizer
R\'enyi non-classicality we characterize analytically.
Our results demonstrate how spatially extended quantum
measurements can be synthesized from local operations
through causal dynamics, while identifying time and the
structure of the automaton as resources for their
implementation. The architecture of the implementation, based on local quantum instruments, guarantees no conflict with the causal structure imposed by locality of the QCA rule.

The paper is organised as follows. In Section~II, we
introduce the problem of local measurements in quantum field theory, together with the algebraic framework of quantum cellular
automata and their causal structure. In Section III, we
formulate local quantum operations and define Operational
Measurement Protocols, establishing their causal properties.
In Section IV, we construct an explicit measurement
protocol and analyse the entanglement and stabilizer
non-classicality of the induced measurements as a function of the internal parameters defining the QCA evolution. Finally,
in Section V, we discuss the implications of our results
and possible future developments.

\section{Impossible Measurements and Quantum Cellular Automata}

The theoretical apparatus used to treat quantum cellular automata is the same as in the algebraic formulation of quantum field theory \cite{HaagKastler1964}. Here, locality of observables is embodied in a net $\Omega \mapsto \alg A_\Omega$, which assigns to each bounded open spacetime region \(\Omega\) a von Neumann algebra \(\alg A_\Omega\) of operators localized in that region. Relativistic causality is expressed kinematically through the axiom of
\emph{microcausality}: if \(\Omega_A\) and \(\Omega_B\) are spacelike
separated, then their local algebras commute, or equivalently $\alg A_{\Omega_A}\subseteq \alg A_{\Omega_B}'$. Microcausality therefore ensures that observables associated with causally disjoint regions are mutually compatible and that local operations performed within one region cannot affect the unconditional statistics of observables localized in the other. However, it does not specify, by itself, how a physical intervention localized in one region is allowed to \emph{transform} observables associated with other regions. This distinction is precisely what makes relativistic measurement theory subtle. While microcausality is a condition on the kinematical organization of observables, a measurement introduces a dynamical transformation. The relevant question is whether the
operations representing physical measurement procedures respect the causal
structure imposed by the theory. A mathematically well-defined local operation need not automatically define a physically admissible procedure in a relativistic setting.

In quantum field theory, this conceptual problem is accompanied by further
algebraic difficulties. Local algebras associated with bounded spacetime
regions are generically type-III von Neumann factors
\cite{Driessler1977}, so the familiar finite-dimensional description of
measurements in terms of density matrices, minimal projections, and
trace-rules cannot be transferred directly to the local
theory. Different approaches have therefore been developed to formulate
measurement processes in a relativistically consistent way. Early proposals
focused on covariant prescriptions for state reduction
\cite{HellwigKraus1970,AharonovAlbert1981}, whereas more recent operational
frameworks describe measurements through localized couplings between the
field and auxiliary probe systems
\cite{FewsterVerch2020,BostelmannFewsterRuep2021,Perche2024,
PranziniKeskiVakkuri2025}. These approaches differ in their formulation,
but they point to the same structural issue: relativistic consistency
cannot be inferred from microcausality alone; one must also control the
causal behaviour of the operations induced by the measurement process.

\subsection{{A Three-Region Scenario}}
\label{subsec:three-region-scenario}

The operational content of the local measurement problem can be isolated
in a three-region configuration
\cite{Sorkin1993,BorstenJubbKells2021}. Let
\(\Omega_A\) and \(\Omega_B\) be spacelike separated spacetime regions, and
let \(\Omega_C\) be an extended intermediate region. By construction,
there is no causal relation allowing information to propagate directly
from \(\Omega_A\) to \(\Omega_B\). Nevertheless, different portions of
\(\Omega_C\) may be causally related to the two regions: one part of
\(\Omega_C\) may lie in the causal future of \(\Omega_A\), while another
part may lie in the causal past of \(\Omega_B\). The potential obstruction arises when a single operation localized over
the whole of \(\Omega_C\) correlates these otherwise causally separated
portions. 

Consider the following sequence of interventions: (i) an agent in \(\Omega_A\) performs a deterministic localized operation, represented in the Heisenberg picture by a unital completely positive map \(\mathcal O_A\), (ii) an agent in \(\Omega_C\) performs a measurement described by a quantum instrument $\{\mathcal J_C^m\}_{m\in\mathbb O}$, (iii) an agent in \(\Omega_B\) measures an observable $O_B\in\alg A_{\Omega_B}$. Let $\overline{\mathcal J}_C := \sum_{m\in\mathbb O}\mathcal J_C^m$ denote the non-selective transformation associated with the intermediate measurement. Starting from an initial normal state \(\omega\), the final expectation value of \(O_B\) is $(\omega \circ  \mathcal O_A \circ \overline{\mathcal J}_C)(O_B)$. Relativistic causality requires this quantity to be independent of the operation performed in \(\Omega_A\). Thus, for any two deterministic operations
\(\mathcal O_A^{(1)}\) and \(\mathcal O_A^{(2)}\) localized in
\(\Omega_A\),
\[
    (\omega \circ  \mathcal O_A^{(1)} \circ \overline{\mathcal J}_C)(O_B) = (\omega \circ  \mathcal O_A^{(2)} \circ \overline{\mathcal J}_C)(O_B)
\] must hold for every state \(\omega\) and every \(O_B\in\alg A_{\Omega_B}\). The crucial object is therefore not the original observable \(O_B\), but
its Heisenberg pullback through the intermediate operation, $\overline{\mathcal J}_C(O_B)$. Although microcausality guarantees $[O_B,X_A]=0$ for every $X_A\in\alg A_{\Omega_A}$, it imposes no analogous condition \emph{a priori} on the transformed
observable. Causal consistency requires
\begin{equation}
    \left[ \ 
        \overline{\mathcal J}_C(O_B),
        X_A
    \ \right]
    =
    0
    \qquad
    \forall\,
    O_B\in\alg A_{\Omega_B},
    \;
    X_A\in\alg A_{\Omega_A}.
    \label{eq:aqft-nosignalling}
\end{equation} Under the assumptions discussed in
\cite[Sec.~III, Claim~1]{BorstenJubbKells2021}, this condition
characterizes precisely the absence of signalling in the corresponding
three-region protocol.

\begin{figure}[t]
\centering

\resizebox{\columnwidth}{!}{%
\begin{tikzpicture}[
    x=1cm,
    y=1cm,
    line cap=round,
    line join=round
]


\def\regionheight{0.52}


\filldraw[
    fill=white,
    draw=black,
    line width=0.9pt
]
    (1.20,1.05) rectangle (4.40,1.57);

\filldraw[
    fill=orange!35,
    draw=black,
    line width=0.9pt
]
    (4.00,2.70) rectangle (7.20,3.22);

\filldraw[
    fill=white,
    draw=black,
    line width=0.9pt
]
    (6.80,4.10) rectangle (10.00,4.62);



\filldraw[
    fill=yellow!25,
    fill opacity=0.40,
    draw=orange!90!red,
    line width=0.85pt
]
    (0.15,5.15) --
    (5.45,5.15) --
    (4.40,1.05) --
    (1.20,1.05) -- cycle;


\filldraw[
    fill=yellow!25,
    fill opacity=0.40,
    draw=orange!90!red,
    line width=0.85pt
]
    (6.80,4.62) --
    (10.00,4.62) --
    (11.05,0.52) --
    (5.75,0.52) -- cycle;


\node[font=\large]
    at (2.80,1.31)
    {$\Omega_A$};

\node[font=\large]
    at (5.60,2.96)
    {$\Omega_C$};

\node[font=\large]
    at (8.40,4.36)
    {$\Omega_B$};


\draw[
    black,
    line width=0.85pt,
    ->,
    >=stealth
]
    (0,0) -- (11.55,0)
    node[below left,font=\large] {$x$};

\draw[
    black,
    line width=0.85pt,
    ->,
    >=stealth
]
    (0,-0.15) -- (0,5.75)
    node[above left,font=\large] {$t$};

\end{tikzpicture}%
}

\caption{
Three-region spacetime configuration.
}
\label{fig:three_regions}

\end{figure}
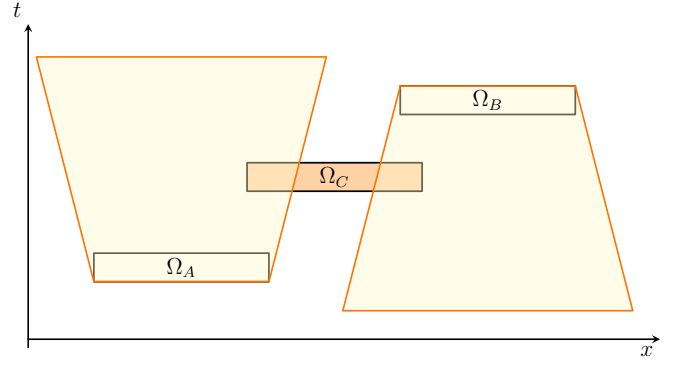

If Equation~\eqref{eq:aqft-nosignalling} holds, the observable induced in
the past by the intermediate measurement remains causally inaccessible to
operations localized in \(\Omega_A\), and the statistics measured in
\(\Omega_B\) cannot depend on the choice of intervention in
\(\Omega_A\). If it fails, the intermediate operation has transformed an
observable originally localized in \(\Omega_B\) into one that is sensitive
to operations in \(\Omega_A\). The measurement performed in
\(\Omega_C\) can then mediate an effective influence between two regions
that were originally spacelike separated. This is the mechanism underlying the \emph{impossible measurement} phenomenon identified by Sorkin and subsequently analysed in algebraic terms \cite{Sorkin1993,Benincasa2014,BorstenJubbKells2021,Jubb2022}. The
pathology is therefore not a failure of microcausality of the original observable net. Rather, it is a failure of the measurement-induced transformation to preserve the causal localization required by Special Relativity. 

Finally, observe that the three-region pathology necessarily requires the intermediate measurement region to have a non-trivial spacetime extension. Indeed, an idealized pointlike intervention cannot simultaneously receive a causal influence from \(\Omega_A\) and transmit one to \(\Omega_B\): if this were possible,
causal transitivity would provide a causal connection from
\(\Omega_A\) to \(\Omega_B\), contradicting their spacelike separation. More generally, the same obstruction is absent whenever the localization region of the intermediate operation does not contain distinct portions having the two required causal relations. The pathology can arise only when \(\Omega_C\) contains one portion that can be influenced from
\(\Omega_A\) and a distinct portion that can in turn influence
\(\Omega_B\). A single operation supported on the whole of
\(\Omega_C\) may then couple these causally distinct portions and thereby
create the effective signalling path characteristic of the three-region
scenario. Thus, it is the causal extension of the intermediate operation,
rather than the mere presence of a local measurement, that makes the
obstruction possible.

\subsubsection{{Why Quantum Cellular Automata?}}
\label{subsec:why-qca}

Quantum Cellular Automata provide a particularly controlled setting in
which the same structural problem can be studied without the additional
analytic complications of relativistic quantum field theory. A QCA is a
discrete-time, discrete-space quantum many-body system whose local
observables form a quasi-local \(C^*\)-algebra and whose dynamics is given
by a locality-preserving automorphism. In the Heisenberg picture, the
automaton therefore acts directly on observables and propagates their
supports within a finite neighbourhood
\cite{SchumacherWerner2004,GrossNesmeVogtsWerner2012}.

The relevance of QCAs for local measurement theory is not that they
reproduce all structures of a relativistic quantum field theory. They do
not: spacetime is discrete, local algebras are finite-dimensional, and no
Lorentz symmetry is assumed. Rather, they retain precisely the algebraic
and operational ingredients that enter the problem of local measurements described
above. First, both frameworks possess a net of local observable algebras embedded
into a quasi-local algebra. For a QCA on \(\mathbb Z^D\), each finite
region $\Omega\Subset\mathbb Z^D$ is associated with a finite-dimensional local algebra $\alg A_\Omega$ and disjoint regions correspond to mutually commuting subalgebras. Localization can therefore be formulated directly at the algebraic level, in close analogy with the net structure of AQFT. Second, QCA dynamics ensures finite speed for propagation of information across the lattice. If an observable is initially localized in a region \(\Omega\), its Heisenberg evolution can spread only inside the causal neighbourhood generated by the automaton. The analogue of a relativistic causal cone is therefore encoded algebraically by the support propagation of the QCA automorphism. This makes it possible to ask with complete precision whether the pullback of an observable, or of an effective measurement effect, remains inside the region allowed by the
causal structure imposed by the QCA itself. Third, the local algebras are finite-dimensional type-I factors. Local quantum operations can consequently be represented directly by completely positive maps, quantum instruments admit ordinary Kraus representations, and measurement effects are genuine finite-dimensional POVM elements. Thus the usual operational language of quantum information can be used without abandoning the algebraic notion of localization. These features make QCAs a minimal operator-algebraic laboratory for
separating the essential causal structure of local measurement theory from the specifically algebraic-theoretic complications of AQFT.

\subsection{Quantum Cellular Automata}
\label{sec:qca}

Quantum Cellular Automata (QCAs) provide a discrete-time, local, and reversible model of quantum dynamics on a lattice of finite-dimensional quantum systems \cite{SchumacherWerner2004,ArrighiNesmeWerner2011}. In the present work they are not used merely as lattice regularizations of continuum quantum field theories. Rather, they provide an autonomous operational setting in which locality, causality, and measurement processes can be formulated directly at the algebraic level. Their relevance for the measurement problem discussed above is twofold. First, the dynamics has strict finite propagation speed: an observable initially localized in a finite region can spread only inside a finite causal neighbourhood after one time step. Second, the algebras associated with bounded regions are finite-dimensional type-I factors. Hence completely positive (CP) maps, Kraus decompositions, quantum instruments, and POVMs are intrinsically well-defined on local subsystems, without the algebraic obstructions associated with type-III local algebras in continuum AQFT \cite{OkamuraOzawa2016}.

\smallskip

Let the spatial lattice be $\mathbb Z^D$, with $D\geq 1$. To each site $x\in\mathbb Z^D$ we associate a finite-dimensional Hilbert space $\mathcal H_x\simeq\mathbb C^d$, with $d < \infty$. For every finite region $\Omega\Subset\mathbb Z^D$, we define
\[
    \mathcal H_\Omega
    :=
    \bigotimes_{x\in\Omega}\mathcal H_x,
    \qquad
    \alg A_\Omega
    :=
    \alg B(\mathcal H_\Omega),
\] where $\alg B(\mathcal H_\Omega)$ denotes the $C^*$-algebra of bounded linear operators on $\mathcal H_\Omega$. Since $\mathcal H_\Omega$ is finite-dimensional, one has $\alg A_\Omega \simeq \alg{Mat}_{d^{|\Omega|}}(\mathbb C)$. If $\Omega_1\subseteq\Omega_2$, we identify $\alg A_{\Omega_1}$ with the subalgebra of $\alg A_{\Omega_2}$ obtained by tensoring with the identity on $\Omega_2\setminus\Omega_1$, that is,
\[
    A_{\Omega_1}
    \mapsto
    A_{\Omega_1}\otimes\mathbf 1_{\Omega_2\setminus\Omega_1}.
\] The assignment $\Omega\mapsto\alg A_\Omega$ defines a net of finite-dimensional local algebras on the spatial lattice. The uniform norm completion of the direct union of all finite local algebras defines the quasi-local algebra of the lattice \cite{RichterWerner1996,SchumacherWerner2004},
\[
    \alg A
    :=
    \overline{\bigcup_{\Omega\Subset\mathbb Z^D}\alg A_\Omega.}^{\|\cdot\|}
\] For each lattice vector $v\in\mathbb Z^D$, let
\[
    \Omega+v:=\{x+v:x\in\Omega\}.
\] The canonical identification
$\mathcal H_\Omega\simeq\mathcal H_{\Omega+v}$ induces a
$*$-isomorphism $\QCA{t}_a|_\Omega:\alg A_\Omega\to\alg A_{\Omega+v}$. These maps extend uniquely to $*$-automorphisms of the quasi-local algebra 
$\QCA{t}_a\in\operatorname{Aut}(\alg A)$, called \emph{shifts}. The map $a\mapsto\QCA{t}_a$ is a representation of the translation group $\mathbb Z^D$ by automorphisms of the quasi-local algebra, indeed,
\[
    \QCA{t}_a\circ\QCA{t}_b=\QCA{t}_{a+b},
    \qquad
    \QCA{t}_0=\operatorname{id}_{\alg A}. \quad \
\] 

A \emph{neighbourhood scheme} is a finite subset
$N\Subset\mathbb Z^D$. Its translates specify the sites supporting the one-step evolution of single-site observables. In particular, the neighbourhood of a site $x\in\mathbb Z^D$ is the translated set
\[
    N(x):=x+N=\{x+n:\ n\in N\}.
\] More generally, for every finite region
$\Omega\Subset\mathbb Z^D$, the neighbourhood scheme induces the
thickening
\[
    N(\Omega)
    :=
    \Omega+N
    :=
    \{x+n:\ x\in\Omega,\ n\in N\}.
\] Equivalently, $N(\Omega)$ is the union of the neighbourhoods of all sites
belonging to $\Omega$. For a given QCA, we take its neighbourhood scheme to be minimal with respect to set inclusion among all finite neighbourhood schemes satisfying the causality condition below.

\begin{definition}
\label{def:qca}
A quantum cellular automaton \emph{(QCA)} on $\mathbb Z^D$ is a $*$-automorphism $\QCA{q} \in\operatorname{Aut}(\alg A)$ such that there exists a finite neighbourhood scheme $N_{\QCA{q}}\Subset\mathbb Z^D$ satisfying
\begin{equation}
    \QCA{q}(\alg A_\Omega)
    \subseteq
    \alg A_{N_{\QCA{q}}(\Omega)}
    \quad
    \text{for every finite} \;
    \Omega\Subset\mathbb Z^D ,
    \label{eq:qca-causality}
\end{equation} while for every strict subset $M \subset N_{\QCA{q}} \Subset \mathbb{Z}^D$ there exists $\Omega\Subset\mathbb Z^d$ such that $\QCA{q}(\alg A_\Omega) \not\subseteq \alg A_{M(\Omega)}$ If, in addition,
\begin{equation}
     \QCA{q}\circ\QCA{t}_a=\QCA{t}_a\circ \QCA{q}
    \qquad
    \forall a\in\mathbb Z^D,
    \label{eq:translation-covariance}
\end{equation} we say that the QCA is translation-covariant.
\end{definition}

Condition \eqref{eq:qca-causality} is the algebraic expression of
finite propagation speed of information across the lattice \cite{SchumacherWerner2004,GrossNesmeVogtsWerner2012}. For a QCA $\QCA{q}$ with neighbourhood scheme $N_{\QCA{q}}$, we write
\[
    N_{\QCA{q}}^0(\Omega):=\Omega,
    \quad
    N_{\QCA{q}}^{t+1}(\Omega)
    :=
    \bigl(
        N_{\QCA{q}}\circ N_{\QCA{q}}^t
    \bigr)(\Omega),
    \qquad
    t\in\mathbb N.
\] By induction,
\begin{equation}
    \QCA{q}^t(\alg A_\Omega)
    \subseteq
    \alg A_{N_{\QCA{q}}^t(\Omega)}
    \qquad
    t\in\mathbb N.
    \label{eq:iterated-causality}
\end{equation} Throughout the paper we adopt the Heisenberg convention according to
which, if $\omega_t$ is the state at time $t$, then
\[
    \omega_{t+1}(A)
    =
    \bigl(\omega_t \circ \QCA{q}\bigr)(A), \qquad A \in \mathsf{A}
\] Accordingly, $\QCA{q}(A)$ is the representative at time $t$ of an
observable $A$ measured at time $t+1$. More generally, $\QCA{q}^t(A)$ is the representative $t$ time steps earlier of the same observable. Hence $N_{\QCA{q}}^t(\Omega)$ bounds the spatial support of the backward Heisenberg propagation of an observable localized in $\Omega$ after $t$ time steps. In this sense, the neighbourhood scheme determines the discrete causal structure of the
automaton \cite[Sec.~II]{SchumacherWerner2004}.

\begin{figure}[h!]
\centering

\resizebox{0.85\columnwidth}{!}{%
\begin{tikzpicture}[
    x=1cm,
    y=1cm,
    line cap=round,
    line join=round
]


\def\cell{0.75}
\def\xstart{0.8}
\def\ylow{0}
\def\yhigh{1.9}
\def\height{0.75}


\fill[red!18]
    (\xstart+2*\cell,\yhigh)
    --
    (\xstart+3*\cell,\yhigh)
    --
    (\xstart+4*\cell,\ylow+\height)
    --
    (\xstart+1*\cell,\ylow+\height)
    -- cycle;


\fill[orange!30]
    (\xstart+1*\cell,\ylow)
    rectangle
    (\xstart+4*\cell,\ylow+\height);

\fill[orange!75]
    (\xstart+2*\cell,\yhigh)
    rectangle
    (\xstart+3*\cell,\yhigh+\height);


\foreach \i in {0,...,5} {

    \draw[line width=0.8pt]
        (\xstart+\i*\cell,\ylow)
        --
        (\xstart+\i*\cell,\ylow+\height);

    \draw[line width=0.8pt]
        (\xstart+\i*\cell,\yhigh)
        --
        (\xstart+\i*\cell,\yhigh+\height);
}


\foreach \y in {
    \ylow,
    \ylow+\height,
    \yhigh,
    \yhigh+\height
} {
    \draw[line width=0.9pt]
        (\xstart-0.20,\y)
        --
        (\xstart+5*\cell+0.20,\y);
}


\draw[
    red!70,
    line width=1.15pt
]
    (\xstart+2*\cell,\yhigh)
    --
    (\xstart+1*\cell,\ylow+\height);

\draw[
    red!70,
    line width=1.15pt
]
    (\xstart+3*\cell,\yhigh)
    --
    (\xstart+4*\cell,\ylow+\height);


\node[font=\large]
    at (\xstart-0.75,\ylow+0.5*\height)
    {$t$};

\node[font=\large]
    at (\xstart-0.75,\yhigh+0.5*\height)
    {$t+1$};

\node[font=\large]
    at (\xstart+2.5*\cell,\ylow-0.42)
    {$\mathsf A_{N_{\mathbf q}(\Omega)}$};

\node[font=\large]
    at (\xstart+2.5*\cell,\yhigh+\height+0.42)
    {$\mathsf A_{\Omega}$};

\node[font=\Large]
    at (\xstart+2.5*\cell,1.43)
    {$\mathbf q$};

\end{tikzpicture}%
}

\caption{
Heisenberg propagation of a local
observable algebra $\mathsf A_\Omega$ under
the QCA automorphism $\mathbf q$.
}
\label{fig:qca_causal_neighbourhood}

\end{figure}
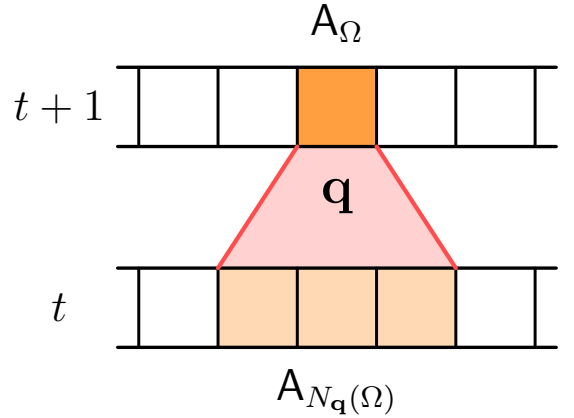

Since $\alg A_\Omega$ and $\alg A_{N_{\QCA{q}}(\Omega)}$ are full matrix algebras and $\alg A_\Omega$ embeds canonically into $\alg A_{N_{\QCA{q}}(\Omega)}$, the representation theory of full matrix algebras, equivalently the finite-dimensional
Skolem--Noether property \cite[Appendix~A, Proposition~12]{SchumacherWerner2004}, implies that the restriction
\[
    \QCA{q}|_{\alg A_\Omega}:
    \alg A_\Omega
    \to
    \alg A_{N_{\QCA{q}}(\Omega)}
\] is unitarily equivalent to the canonical inclusion $A \mapsto A\otimes \mathbf 1_{N_{\QCA{q}}(\Omega)\setminus\Omega}$. Thus there exists a unitary $U_{N_{\QCA{q}}(\Omega)} \in \alg A_{N_{\QCA{q}}(\Omega)}$ such that, for every $A\in\alg A_\Omega$,
\begin{equation}
    \QCA{q}(A)
    =
    U_{N_{\QCA{q}}(\Omega)}^\dagger
    \bigl(
        A\otimes
        \mathbf 1_{N_{\QCA{q}}(\Omega)\setminus\Omega}
    \bigr)
    U_{N_{\QCA{q}}(\Omega)}.
    \label{eq:local-implementation}
\end{equation} The implementing unitary is not canonical. If
$U_{N_{\QCA{q}}(\Omega)}$ is one choice, then $VU_{N_{\QCA{q}}(\Omega)}$ implements the same restriction whenever $V$ is a unitary belonging to the relative commutant of the canonically embedded algebra, \emph{i.e.},
\[
    V
    \in
    \bigl(
        \alg A_\Omega\otimes
        \mathbf 1_{N_{\QCA{q}}(\Omega)\setminus\Omega}
    \bigr)'
    \cap
    \alg A_{N_{\QCA{q}}(\Omega)} = \mathbf{1}_\Omega\otimes
        \mathsf{A}_{N_{\QCA{q}}(\Omega)\setminus\Omega}.
\] The QCA is therefore the global automorphism $\QCA{q}$, not a
distinguished family of local implementing unitaries.

\smallskip

For our purposes, it is useful to regard a QCA as defining a net of local algebras on the discrete spacetime lattice \(\mathbb Z^D\times\mathbb Z\). We choose \(t=0\) as a common reference time slice. For every \(t\in\mathbb Z\) and every finite spatial region \(\Omega\Subset\mathbb Z^D\), we define
\begin{equation}
    \alg A_{(\Omega,t)}
    :=
    \QCA{q}^{t}(\alg A_\Omega).
    \label{eq:spacetime-net}
\end{equation} According to the Heisenberg convention introduced above,
\(\alg A_{(\Omega,t)}\) is therefore the algebra of observables localized in
\(\Omega\) at time \(t\), represented on the reference time slice \(t=0\). Notice that the label \(\Omega\) refers to the localization region at time \(t\), and not, in general, to the spatial support of its representative on the reference slice.

At every fixed time \(t\), the assignment $\Omega\mapsto\alg A_{(\Omega,t)}$ satisfies \emph{isotony}:
\[
    \Omega_1\subseteq\Omega_2
    \quad\Longrightarrow\quad
    \alg A_{(\Omega_1,t)}
    \subseteq
    \alg A_{(\Omega_2,t)}.
\] Moreover, since \(\QCA{q}^{t}\) is a \( * \)-automorphism, equal-time
\emph{locality} is preserved:
\[
    \Omega_1\cap\Omega_2=\varnothing
    \quad\Longrightarrow\quad
    \bigl[
        \alg A_{(\Omega_1,t)},
        \alg A_{(\Omega_2,t)}
    \bigr]
    =0.
\] The neighbourhood scheme determines the causal relation between different
time slices. Let \(t_2>t_1\), and let \(\Omega_1,\Omega_2\Subset\mathbb Z^D\). By iterated causality, an observable localized in \(\Omega_2\) at time \(t_2\), when represented \(t_2-t_1\) time steps earlier on the slice \(t_1\), belongs to
\[
    \QCA{q}^{\,t_2-t_1}(\alg A_{\Omega_2})
    \subseteq
    \alg A_{N_{\QCA{q}}^{\,t_2-t_1}(\Omega_2)}.
\] Thus \(N_{\QCA{q}}^{\,t_2-t_1}(\Omega_2)\) contains the spatial support on the earlier slice \(t_1\) of the backward Heisenberg propagation of observables localized in \(\Omega_2\) at time \(t_2\). We therefore say that the spacetime regions \((\Omega_1,t_1)\) and \((\Omega_2,t_2)\), with \(t_2>t_1\), are
\emph{causally disjoint} if
\begin{equation}
    \Omega_1
    \cap
    N_{\QCA{q}}^{\,t_2-t_1}(\Omega_2)
    =
    \varnothing.
    \label{eq:causal-disjointness}
\end{equation} Equivalently, the backward Heisenberg causal cone of the later region
\((\Omega_2,t_2)\) does not intersect the earlier region \((\Omega_1,t_1)\). This causal disjointness condition implies a discrete analogue of
microcausality. Indeed, let
\(A_1\in\alg A_{\Omega_1}\) and
\(A_2\in\alg A_{\Omega_2}\). Since \(t_2>t_1\),
\[
\begin{aligned}
    \bigl[
        \QCA{q}^{t_1}(A_1),
        \QCA{q}^{t_2}(A_2)
    \bigr]
    &=
    \QCA{q}^{t_1}
    \left(
        \bigl[
            A_1,
            \QCA{q}^{\,t_2-t_1}(A_2)
        \bigr]
    \right).
\end{aligned}
\] By iterated causality,
\[
    \QCA{q}^{\,t_2-t_1}(A_2)
    \in
    \alg A_{N_{\QCA{q}}^{\,t_2-t_1}(\Omega_2)}.
\] Hence, if Equation~\eqref{eq:causal-disjointness} holds, locality of the
quasi-local algebra on the slice \(t_1\) gives
\[
    \bigl[
        A_1,
        \QCA{q}^{\,t_2-t_1}(A_2)
    \bigr]
    =
    0.
\] Since this holds for arbitrary
\(A_1\in\alg A_{\Omega_1}\) and
\(A_2\in\alg A_{\Omega_2}\),
\begin{equation}
    \left[
        \alg A_{(\Omega_1,t_1)},
        \alg A_{(\Omega_2,t_2)}
    \right]
    =
    0.
    \label{eq:spacetime-locality}
\end{equation} For \(t_1>t_2\), the same definition is applied after exchanging the roles of the earlier and later spacetime regions.

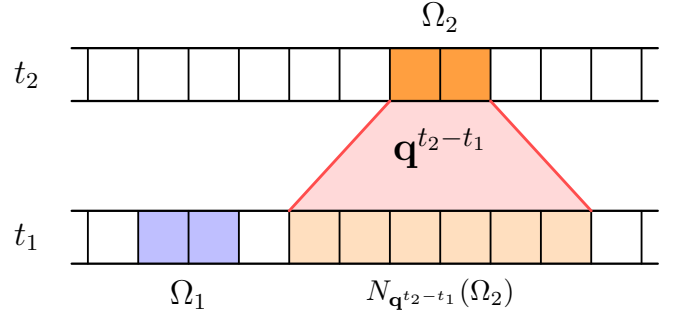
\begin{figure}[h!]
\centering

\resizebox{\columnwidth}{!}{%
\begin{tikzpicture}[
    x=1cm,
    y=1cm,
    line cap=round,
    line join=round
]


\def\cell{0.62}
\def\xstart{0.8}
\def\ylow{0}
\def\yhigh{2.0}
\def\height{0.65}


\fill[red!15]
    (\xstart+6*\cell,\yhigh)
    --
    (\xstart+8*\cell,\yhigh)
    --
    (\xstart+10*\cell,\ylow+\height)
    --
    (\xstart+4*\cell,\ylow+\height)
    -- cycle;


\fill[blue!25]
    (\xstart+1*\cell,\ylow)
    rectangle
    (\xstart+3*\cell,\ylow+\height);

\fill[orange!25]
    (\xstart+4*\cell,\ylow)
    rectangle
    (\xstart+10*\cell,\ylow+\height);

\fill[orange!75]
    (\xstart+6*\cell,\yhigh)
    rectangle
    (\xstart+8*\cell,\yhigh+\height);


\foreach \i in {0,...,11} {

    \draw[line width=0.65pt]
        (\xstart+\i*\cell,\ylow)
        --
        (\xstart+\i*\cell,\ylow+\height);

    \draw[line width=0.65pt]
        (\xstart+\i*\cell,\yhigh)
        --
        (\xstart+\i*\cell,\yhigh+\height);
}

\foreach \y in {
    \ylow,
    \ylow+\height,
    \yhigh,
    \yhigh+\height
} {
    \draw[line width=0.8pt]
        (\xstart-0.20,\y)
        --
        (\xstart+11*\cell+0.20,\y);
}


\draw[
    red!70,
    line width=1pt
]
    (\xstart+6*\cell,\yhigh)
    --
    (\xstart+4*\cell,\ylow+\height);

\draw[
    red!70,
    line width=1pt
]
    (\xstart+8*\cell,\yhigh)
    --
    (\xstart+10*\cell,\ylow+\height);


\node[font=\large]
    at (\xstart-0.75,\ylow+0.5*\height)
    {$t_1$};

\node[font=\large]
    at (\xstart-0.75,\yhigh+0.5*\height)
    {$t_2$};

\node[font=\large]
    at (\xstart+2*\cell,\ylow-0.38)
    {$\Omega_1$};

\node[font=\large]
    at (\xstart+7*\cell,\yhigh+\height+0.38)
    {$\Omega_2$};

\node[font=\normalsize]
    at (\xstart+7*\cell,\ylow-0.38)
    {$N_{\mathbf q^{t_2-t_1}}(\Omega_2)$};

\node[font=\Large]
    at (\xstart+7*\cell,1.42)
    {$\mathbf q^{t_2-t_1}$};

\end{tikzpicture}%
}

\caption{
Causally disjoint spacetime regions
$(\Omega_1,t_1)$ and $(\Omega_2,t_2)$,
with $t_1<t_2$. The causal neighbourhood
of $\Omega_2$ does not intersect $\Omega_1$.
}
\label{fig:causal_disjointness}

\end{figure}

Thus a QCA provides an algebraic discrete-spacetime analogue of the
kinematical part of an AQFT: it consists of a net of local algebras
together with a causal time evolution. The crucial difference is that
the local algebras are finite-dimensional type-I factors rather than
type-III von Neumann algebras. This is precisely what allows us to
formulate local measurement operations directly as completely
positive instruments on bounded regions. In the following section we
use this structure to define localized quantum operations and
measurement protocols.

\smallskip

We now restrict attention to one-dimensional, translation-covariant QCAs with trivial index, \emph{i.e.}, $\operatorname{ind}(\QCA{q})=1$ \cite{GrossNesmeVogtsWerner2012}. For a one-dimensional QCA, the index $\operatorname{ind}(\QCA{q})\in\mathbb{Q}_{>0}$ is a multiplicative invariant that quantifies the net flow of quantum information across a spatial cut. The value \(\operatorname{ind}(\QCA{q})=1\) characterizes the absence of a net transport of quantum information across the chain. A fundamental result of Gross, Nesme, Vogts, and Werner states that a one-dimensional QCA has index one if and only if it is locally implementable by a finite-depth quantum circuit (FDQC) with uniformly bounded local gates \cite[Sec.~7.3, Theorem~9]{GrossNesmeVogtsWerner2012}. For a QCA with arbitrary finite propagation range, the circuit can be expressed, after grouping finitely many adjacent cells into larger cells, in a two-layer form. The structure of such finite-depth decompositions is further analysed for one-dimensional quantum cellular automata in \cite{vanrietvelde2025causaldecompositions1dquantum}. In the qubit case, the relevant normal form can be obtained from the classification of Schumacher and Werner, specifically from Sec.~IV.7, Proposition~10, and Eq.~(37) of \cite{SchumacherWerner2004}. The following result gives the corresponding normal form on the nearest-neighbour qubit chain.

\begin{theorem}
\label{thm:qubit-qca-classification}
Let $\QCA{q}\in\operatorname{Aut}(\alg A)$ be a translation covariant QCA on \(\mathbb{Z}\), with single-site algebras $\alg A_{\{x\}}\simeq\operatorname{Mat}_{2}(\mathbb{C})$, and suppose that its one-step neighbourhood scheme is \(N_{\QCA{q}} = \{-1,0,1\}\), namely
\[
    \QCA{q}\bigl(\alg A_{\{x\}}\bigr)
    \subseteq
    \alg A_{\{x-1,x,x+1\}},
    \qquad
    x\in\mathbb{Z}.
\] Fix the computational basis $\{|0\rangle,|1\rangle\}$ of \(\mathbb{C}^{2}\), used identically at every lattice site. Then $\operatorname{ind}(\QCA{q})=1$ in the sense of \cite{GrossNesmeVogtsWerner2012} if and only if there exist single-qubit unitaries $U,V\in\mathrm{U}(2)$ and a phase $\varphi\in(-\pi,\pi]$ such that
\begin{equation}
    \QCA{q}
    =
    \QCA{v}\circ
    \QCA{c}_{\varphi}\circ
    \QCA{u}.
    \label{eq:qubit-qca-normal-form}
\end{equation} Here \(\QCA{u}\) and \(\QCA{v}\) denote the translation-invariant automorphisms induced by the single-qubit unitaries
\(U,V\in\mathrm{U}(2)\): for every finite region
\(\Omega\Subset\mathbb{Z}\) and every \(A\in\alg A_{\Omega}\),
\begin{align}
    \QCA{u}(A)
    &:=
    (
        \bigotimes_{x\in\Omega}U_{\{x\}}
    )^{\dagger}
    \, A \, 
    (
        \bigotimes_{x\in\Omega}U_{\{x\}}
    ),\\
    \QCA{v}(A)
    &:=
    (
        \bigotimes_{x\in\Omega}V_{\{x\}}
    )^{\dagger}
    \, A \, 
    (
        \bigotimes_{x\in\Omega}V_{\{x\}}
    ).
\end{align} The automorphism \(\QCA{c}_{\varphi}\) is generated by commuting nearest-neighbour controlled-\(Z\)-rotation gates
\[
    \operatorname{CR}_{Z}(\varphi)
    :=
    \sum_{j\in\mathbb{Z}_2}
    |j\rangle\langle j|
    \otimes
    \exp\!\left(
        -\mathrm{i}\frac{\varphi}{2}jZ
    \right),
\] written in the ordered basis
\(\{|00\rangle,|01\rangle,|10\rangle,|11\rangle\}\). Its action on a
single-site observable \(X_{\{x\}}\in\alg A_{\{x\}}\) is
\begin{equation}
    \QCA{c}_{\varphi}
    \bigl(X_{\{x\}}\bigr)
    =
    A_{\varphi}(x)^{\dagger}
    X_{\{x\}}
    A_{\varphi}(x),
    \label{eq:controlled-phase-local-rule}
\end{equation} where
\[
    A_{\varphi}(x)
    :=
    \operatorname{CR}_{Z}(\varphi)_{\{x-1,x\}}
    \operatorname{CR}_{Z}(\varphi)_{\{x,x+1\}}.
\] Since all these gates are diagonal in the computational basis, they commute, and the local rule extends consistently to every local algebra
\(\alg A_{\Omega}\).
\end{theorem}

The controlled-\(Z\)-rotation gate is locally equivalent to the standard controlled-phase gate
\[
    \operatorname{CPZ}(\varphi)
    :=
    \operatorname{diag}
    \left(
        1,1,1,e^{\mathrm{i}\varphi}
    \right)=
    \bigl(D_{\varphi}\otimes\mathbf{1}\bigr)
    \operatorname{CR}(\varphi),
\] where $D_{\varphi} := |0\rangle\langle0| + e^{\mathrm{i}\varphi/2} |1\rangle\langle1|$. Thus, replacing every controlled-\(Z\)-rotation gate by a \(\operatorname{CPZ}(\varphi)\) gate changes the QCA only by translation-invariant on-site phase gates. These factors can be absorbed into the on-site automorphisms \(\QCA{u}\) and \(\QCA{v}\). Consequently, the two gate conventions define the same normal-form class up to local pre- and post-processing.

\smallskip The characterization above shows that the family of QCAs used in our examples in the following sections is not an arbitrary collection of local update rules: it belongs to the structurally classified class of translation-covariant, nearest-neighbour, index-one qubit QCAs.

\section{Local Quantum Operations and Measurement Protocols}
\label{sec:local-operations-omps}

We now formulate measurements as physical operations localized on finite regions of the QCA lattice. Throughout this section, we work in
the Heisenberg picture. A state of the lattice is a positive normalized linear functional on the quasi-local algebra,
\[
\omega:\alg A\to\mathbb{C},
    \qquad
    \omega(A^*A)\geq 0,
    \quad
    \omega(\mathbf{1})=1.
\] For every finite region \(\Omega\Subset\mathbb{Z}^{D}\), its local marginal is the restriction $\omega_{\Omega} := \omega\big|_{\alg A_{\Omega}}$. Since \(\alg A_{\Omega}\) is finite-dimensional, there exists a unique density matrix \(\rho_{\Omega}\in\alg A_{\Omega}\) such that, for any $A\in\alg A_{\Omega}$,
\[
    \omega(A)
    \equiv \omega_\Omega(A) =
    \operatorname{Tr}_{\Omega}(\rho_{\Omega}A),
    \qquad
    A\in\alg A_{\Omega}.
\] A measurement localized in \(\Omega\) is described by a positive operator-valued measure (POVM)
\[
    \{E_\Omega^i\}_{i\in\mathbb O}
    \subset\alg A_\Omega,
    \qquad
    E_\Omega^i\geq 0,
    \qquad
    \sum_{i\in\mathbb O}E_\Omega^i=\mathbf 1_\Omega.
\] where $\mathbb{O}$ is a finite set of outcomes. The operator \(E_\Omega^i\) is the \emph{local effect} associated with the
outcome \(i\), whose probability in the state \(\omega\) is
\[
    p(i\mid\omega)
    =
    \omega(E_\Omega^i)
    =
    \operatorname{Tr}_\Omega(\rho_\Omega E_\Omega^i).
\] A POVM specifies only the outcome statistics. To describe also the state transformation associated with each outcome, one must introduce
quantum operations and, subsequently, quantum instruments.

\smallskip

Let \(\Omega\Subset\mathbb{Z}^{D}\) be a finite region. A \emph{quantum operation} localized in \(\Omega\) is a completely positive subunital map
\[
    \mathcal{J}_{\Omega}:\alg A_{\Omega}\to\alg A_{\Omega},
    \qquad
    \mathcal{J}_{\Omega}(\mathbf{1}_{\Omega})
    \leq
    \mathbf{1}_{\Omega}.
\] For every finite region
\(\Delta\Subset\mathbb{Z}^{D}\) containing \(\Omega\), there is a canonical identification $\alg A_{\Delta} \simeq \alg A_{\Omega} \otimes \alg A_{\Delta\setminus\Omega}$. The operation is extended trivially outside \(\Omega\) by requiring
\begin{equation}
    \mathcal J\big|_{\alg A_\Delta}
    =
    \mathcal J_\Omega
    \otimes
    \operatorname{id}_{\Delta\setminus\Omega}.
    \label{eq:localized-operation-extension}
\end{equation} These maps are compatible with the canonical embeddings of the local algebras. Since they are uniformly bounded, they define a unique completely positive subunital map $\mathcal J:\alg A\to\alg A$ on the quasi-local algebra.

\smallskip 
The observable $E^{\mathcal{J}} := \mathcal{J}(\mathbf{1})$ is the local effect associated with the operation. Since \(\mathcal{J}\) is subunital, $0\leq E^{\mathcal{J}} \leq\mathbf{1}$. Operationally, it encodes the occurrence probability of the outcome represented by
$\mathcal J$: if the system is prepared in a state \(\omega\), the corresponding probability is
\[
    p(\mathcal{J}\mid\omega)
    :=
    \omega(E^{\mathcal{J}})
    =
    (\omega \circ \mathcal{J})(\mathbf{1}).
\] If \(p(\mathcal{J}\mid\omega)>0\), the normalized conditional state is
\[
    \omega^{\mathcal{J}}
    :=
    \frac{
        \omega \circ \mathcal{J}
    }{
        (\omega \circ \mathcal{J})(\mathbf{1})
    } = \frac{
        \omega \circ \mathcal{J}
    }{
        \omega(E^{\mathcal{J}})
    }.
\] Thus, while the full map $\mathcal J$ specifies both the outcome statistics and the corresponding transformation of the system, the effect $E_{\mathcal J}$ retains only its statistical content. A quantum operation localized in \(\Omega\) is called a local
\emph{quantum channel} if it is unital, namely, $\mathcal C_\Omega(\mathbf 1_\Omega)=\mathbf 1_\Omega$.

\smallskip 

A quantum instrument provides a decomposition of a localized quantum channel into selective branches. More precisely, a finite-outcome \emph{quantum instrument} localized in \(\Omega\) is a family
\begin{equation}
    \texttt{J}_{\Omega}
    =
    \left\{
        \mathcal{J}_{\Omega}^{m}
    \right\}_{m\in \mathbb{O}},
\end{equation} where \(\mathbb{O}\) is a finite outcome set and each $\mathcal{J}_{\Omega}^{m}: \alg A_{\Omega}\to\alg A_{\Omega}$ is a local quantum operation, such that the corresponding non-selective operation
\begin{equation}
    \overline{\mathcal{J}}_{\Omega}
    :=
    \sum_{m\in\mathbb O}
    \mathcal{J}_{\Omega}^{m}
\end{equation} is unital. For each outcome \(m\in\mathbb O\), the associated local effect is $E_{\Omega}^{m} := \mathcal{J}_{\Omega}^{m}(\mathbf{1}_{\Omega})$. The normalization condition is therefore equivalent to $\sum_{m\in\mathbb O}E_{\Omega}^{m} = \mathbf{1}_{\Omega}$, so that $\{ E_\Omega^m \}_{m\in\mathbb O}$ is the POVM associated with the instrument.

\smallskip

Let \(\mathcal J^m\) denote the quasi-local extension of the branch \(\mathcal J_\Omega^m\), obtained by acting trivially outside \(\Omega\). If the system is prepared in a state \(\omega\), the probability of obtaining the outcome \(m\) is
\[
    p(m\mid\omega)
    =
    \omega(E_\Omega^m)
    =
    (\omega\circ\mathcal J^m)(\mathbf 1).
\] Whenever \(p(m\mid\omega)>0\), the corresponding normalized conditional state is
\[
    \omega^m
    =
    \frac{
        \omega\circ\mathcal J^m
    }{
        \omega(E_\Omega^m)
    }.
\] 

The distinction between selective and non-selective operations is particularly relevant for locality. Let \(\Lambda\Subset\mathbb Z^D\) be disjoint from \(\Omega\). Under the
canonical identification $\alg A_{\Omega\cup\Lambda} \simeq \alg A_\Omega\otimes\alg A_\Lambda$ the localized extension of a selective branch satisfies, for every
\(B\in\alg A_\Lambda\),
\[
    \mathcal J^m
    \left(
        \mathbf 1_\Omega\otimes B
    \right)
    =
    E_\Omega^m\otimes B.
\] Consequently,
\[
    \omega^m(B)
    =
    \frac{
        \omega(E_\Omega^m\otimes B)
    }{
        \omega(E_\Omega^m)
    }.
\] In general, this need not coincide with \(\omega(B)\). Conditioning on a particular outcome may therefore change the statistics assigned to
observables localized in \(\Lambda\), because the conditioning can reveal correlations already present in the initial state between \(\Omega\) and
\(\Lambda\). This change in the conditional statistics does not constitute signalling:
access to the state \(\omega^m\) requires conditioning on the outcome \(m\),
and hence classical information about the measurement result. By contrast, when the measurement outcome is ignored, the relevant transformation is the non-selective channel $\overline{\mathcal J}$. Since the instrument is normalized,
\[
    (\omega\circ\overline{\mathcal J})(B)
    =
    \omega(B),
    \qquad
    B\in\alg A_\Lambda.
\] Thus every local quantum instrument leaves the unconditional statistics of observables localized outside its support unchanged, even though its
individual selective branches may induce different conditional statistics on those same observables. Accordingly, operational no-signalling is
expressed by the non-selective channel: without access to the measurement outcome, no statistics observable in a region disjoint from \(\Omega\) can
be affected by the local instrument.

\smallskip

Although the definitions above are formulated directly in terms of
completely positive maps on local algebras, their finite-dimensional
Kraus representations will be useful in the explicit construction of
measurement protocols.

Let \(\Omega\Subset\mathbb Z^D\) be a finite region, and let $\mathcal J_\Omega: \alg A_\Omega\to\alg A_\Omega$ be a local quantum operation. Since \(\alg A_\Omega\) is finite-dimensional, there exists a finite family of operators
\(\{K^k_\Omega\}_k\subset\alg A_\Omega\) such that
\cite{Kraus1971,Choi1975}
\begin{equation}
    \mathcal J_\Omega(A)
    =
    \sum_k
    {K^k_\Omega}^\dagger A K^k_\Omega,
    \qquad
    A\in\alg A_\Omega.
    \label{eq:kraus-local-operation}
\end{equation} The operators \(K^k_\Omega\) are referred to as Kraus operators of
\(\mathcal J_\Omega\); the corresponding Kraus representation is, in
general, not unique. The subunitality condition is equivalent to
\[
    \sum_k
    {K^k_\Omega}^\dagger K^k_\Omega
    \leq
    \mathbf 1_\Omega.
\] Accordingly, the effect associated with the operation is
\begin{equation}
    E_{\mathcal J_\Omega}
    =
    \mathcal J_\Omega(\mathbf 1_\Omega)
    =
    \sum_k
    {K^k_\Omega}^\dagger {K^k_\Omega}.
    \label{eq:kraus-operation-effect}
\end{equation} If \(\mathcal J_\Omega\) is a channel, then it is unital and
\[
    \sum_k
    {K^k_\Omega}^\dagger K^k_\Omega
    =
    \mathbf 1_\Omega.
\] The localized extension of \(\mathcal J_\Omega\) to a finite region
\(\Delta\supseteq\Omega\) is obtained by extending each Kraus operator
trivially outside \(\Omega\): for every \(X\in\alg A_\Delta\),
\begin{equation}
    \mathcal J\big|_{\alg A_\Delta}(X)
    =
    \sum_k
    \left(
        {K^k_\Omega}^\dagger
        \otimes
        \mathbf 1_{\Delta\setminus\Omega}
    \right)
    X
    \left(
        {K^k_\Omega}
        \otimes
        \mathbf 1_{\Delta\setminus\Omega}
    \right).
    \label{eq:kraus-local-operation-extension}
\end{equation} 

Consider now a finite-outcome instrument $\texttt{J}_\Omega = \{ \mathcal J_\Omega^m \}_{m\in\mathbb O}$ localized in \(\Omega\). Each selective branch admits a Kraus
representation
\begin{equation}
    \mathcal J_\Omega^m(A)
    =
    \sum_k
    {K^{(m,k)}_\Omega}^\dagger
    A
    {K^{(m,k)}_\Omega},
    \qquad
    A\in\alg A_\Omega,
    \label{eq:kraus-instrument-branch}
\end{equation}
where the first index \(m\) labels the measurement outcome and the
second index \(k\) labels the Kraus operators belonging to that branch.
The corresponding local effect is given by
\begin{equation}
    E_\Omega^m
    :=
    \mathcal J_\Omega^m(\mathbf 1_\Omega)
    =
    \sum_k
    {K^{(m,k)}_\Omega}^\dagger
    {K^{(m,k)}_\Omega},
    \label{eq:kraus-instrument-effect}
\end{equation} and normalization of the instrument is equivalent to
\[
    \sum_{m\in \mathbb{O}}
    \sum_k
    {K^{(m,k)}_\Omega}^\dagger
    {K^{(m,k)}_\Omega}
    =
    \mathbf 1_\Omega.
\]

\subsection{Operational Measurement Protocols}
\label{subsec:omps}

We now introduce measurement architectures designed to reconstruct the
statistics of effects supported on spatially extended regions of the lattice, including effects that are non-separable across a spatial partition. The elementary interventions are nevertheless required to be strictly localized: each of them acts on a single-site algebra, and simultaneous interventions are implemented only as parallel compositions on single-site tensor factors.

\smallskip 

The spatial extension and the possible quantum correlations of the effective measurement are therefore not introduced through a primitive non-local operation. They arise from the concatenation of strictly local instruments with the causal dynamics generated by the QCA. Starting from a localized readout region \(\Omega_0\), the automorphism \(\QCA{q}\) propagates local effects through the successive causal neighbourhoods $\Omega_{\ell} = N_{\QCA{q}}^{\ell}(\Omega_0)$. This allows an observable initially localized in \(\Omega_0\) to be represented on an earlier time slice by an effective observable supported on the extended region \(\Omega_{\ell}\).

\smallskip 
The architecture is non-adaptive in the following precise sense: the local instruments, their localization regions, and their order of application are fixed before the protocol begins. No external classical message obtained at one site is used to choose an instrument at a causally disjoint site. This does not exclude communication mediated by the physical dynamics of the automaton. Information produced by a local intervention may be encoded in the
local quantum degrees of freedom and subsequently propagated by
\(\QCA{q}\) inside its causal neighbourhood. A later fixed instrument may then respond to those degrees of freedom. Such a mechanism is part of the dynamics of the quantum system and respects the causal structure determined by \(N_{\QCA{q}}\); it is distinct from the adaptive classical communication used in an LOCC protocol
\cite[Sec.~2]{Chitambar2014}. 

\smallskip Accordingly, the relation between OMPs and LOCC should not be understood as a general inclusion or separation result. Rather, the two frameworks describe different implementation architectures. LOCC protocols use local operations whose later choices may depend on an externally communicated classical transcript. OMPs use a fixed sequence of strictly local instruments, while all communication occurring during the protocol is carried by the physical degrees of freedom evolving under the QCA. Any classical comparison of the
recorded outcomes may be performed afterwards, in a common causal future, without entering the implementation of the effective measurement itself.

\begin{definition}
\label{def:instrument-layer}
Let \(\Omega\Subset\mathbb{Z}^{D}\) be a finite region. For every site
\(x\in\Omega\), let $\texttt{J}_{\{x\}} = \{ \mathcal{J}_{\{x\}}^{m_x} \}_{m_x\in \mathbb{O}_x}$ be a finite-outcome quantum instrument localized on the single-site algebra \(\alg A_{\{x\}}\). If no physical intervention is performed at \(x\), we consider
\[
    \mathbb{O}_x=\{\ast\},
    \qquad
    \mathcal{J}_{\{x\}}
    =
    \operatorname{id}_{\{x\}}.
\] The \emph{instrument layer} supported on \(\Omega\) is the quantum instrument $\texttt{J}_{\Omega} = \left\{ \mathcal{J}_{\Omega}^{m} \right\}_{m\in \mathbb{O}_{\Omega}}$ where $\mathbb{O}_{\Omega} := \prod_{x\in\Omega}\mathbb{O}_x$ is the set of joint outcomes $m := (m_x)_{x\in\Omega}$, and the branch associated with \(m\) is the tensor product
\[
    \mathcal{J}_{\Omega}^{m}
    :=
    \bigotimes_{x\in\Omega}
    \mathcal{J}_{\{x\}}^{m_x}.
\]
\end{definition} 

Since the single-site maps act on distinct tensor factors, their composition is order-independent. In addition, since every single-site non-selective instrument is unital, the corresponding non-selective layer map is unital:
\[
    \overline{\mathcal{J}}_{\Omega}
    :=
    \sum_{m\in \mathbb{O}_{\Omega}}
    \mathcal{J}_{\Omega}^{m} = \bigotimes_{x \in \Omega} \; (\sum_{m_x \in \mathbb{O}_x} \mathcal{J}_{\{x\}}^{m_x} ) = \bigotimes_{x \in \Omega} \overline{\mathcal{J}}_{\{x\}}.
\]

\begin{definition}
\label{def:omp}
Let $\QCA{q}\in\operatorname{Aut}(\mathsf{A})$ be a translation-covariant QCA on \(\mathbb{Z}^{D}\), with neighbourhood
scheme $N_{\QCA{q}}$. Let $\Omega_0\Subset\mathbb{Z}^{D}$ be a finite region and define recursively
\[
    \Omega_\ell
    :=
    N_{\QCA{q}}^{\ell}(\Omega_0),
    \qquad
    \ell \in\mathbb{N}.
\] An \emph{Operational Measurement Protocol} (OMP) of order \(n\) rooted at \(\Omega_0\) is the quantum instrument
\[
    \texttt{P}_{\QCA{q}}
    =
    \left\{
        \mathcal{P}_{\QCA{q}}^{m}
    \right\}_{m\in \mathbb{O}_{\texttt{P}}},
\] obtained by concatenating \(n+1\) instrument layers $\texttt{J}_{\Omega_0}, \texttt{J}_{\Omega_1}, \ldots, \texttt{J}_{\Omega_{n}}$ with \(n\) applications of the QCA dynamics. More precisely, for every $\ell=0,\ldots,n$, let $\texttt{J}_{\Omega_\ell} = \{ \mathcal{J}_{\Omega_\ell}^{m_\ell} \}_{m_\ell\in \mathbb{O}_{\Omega_\ell}}$ be an instrument layer supported on \(\Omega_\ell\). The set of complete outcomes is $\mathbb{O}_{\texttt{P}} := \prod_{\ell = 0}^{n}\mathbb{O}_{\Omega_\ell}$ and a complete outcome is denoted by $m= (m_0,\ldots,m_{n}) \in \mathbb{O}_{\texttt{P}}$. The branch associated with the complete outcome \(m\) is the
completely positive map
\begin{equation}
    \mathcal{P}_{\QCA{q}}^{m}
    :=
    \mathcal{J}_{\Omega_{n}}^{m_{n}}
    \circ
    \QCA{q}
    \circ
    \mathcal{J}_{\Omega_{n-1}}^{m_{n-1}}
    \circ
    \QCA{q}
    \circ\cdots\circ
    \QCA{q}
    \circ
    \mathcal{J}_{\Omega_0}^{m_0}.
    \label{eq:omp-branch-map}
\end{equation}
\end{definition}

The non-selective operation associated with the protocol is
\[
    \overline{\mathcal{P}}_{\QCA{q}}
    :=
    \sum_{m\in \mathbb{O}_{\texttt{P}}}
    \mathcal{P}_{\QCA{q}}^{m} = \overline{\mathcal{J}}_{\Omega_n} \circ \QCA{q} \circ \cdots \circ \QCA{q} \circ \overline{\mathcal{J}}_{\Omega_{0}}.
\] Since every layer is a quantum instrument, its non-selective map is unital; since \(\QCA{q}\) is a unital \(*\)-automorphism,
\(\overline{\mathcal{P}}_{\QCA{q}}\) is a quantum channel.

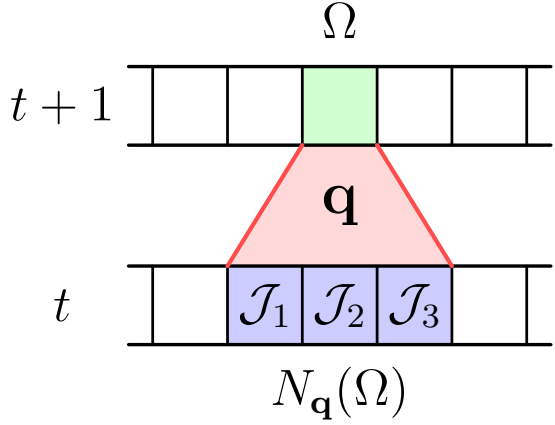
\begin{figure}[t]
\centering

\resizebox{0.85\columnwidth}{!}{%
\begin{tikzpicture}[
    x=1cm,
    y=1cm,
    line cap=round,
    line join=round
]


\def\cell{0.62}
\def\xstart{0.8}
\def\ylow{0}
\def\yhigh{1.65}
\def\height{0.65}


\fill[red!15]
    (\xstart+2*\cell,\yhigh)
    --
    (\xstart+3*\cell,\yhigh)
    --
    (\xstart+4*\cell,\ylow+\height)
    --
    (\xstart+1*\cell,\ylow+\height)
    -- cycle;


\fill[green!18]
    (\xstart+2*\cell,\yhigh)
    rectangle
    (\xstart+3*\cell,\yhigh+\height);

\fill[blue!20]
    (\xstart+1*\cell,\ylow)
    rectangle
    (\xstart+4*\cell,\ylow+\height);


\foreach \i in {0,...,5} {

    \draw[line width=0.65pt]
        (\xstart+\i*\cell,\ylow)
        --
        (\xstart+\i*\cell,\ylow+\height);

    \draw[line width=0.65pt]
        (\xstart+\i*\cell,\yhigh)
        --
        (\xstart+\i*\cell,\yhigh+\height);
}


\foreach \y in {
    \ylow,
    \ylow+\height,
    \yhigh,
    \yhigh+\height
} {
    \draw[line width=0.8pt]
        (\xstart-0.20,\y)
        --
        (\xstart+5*\cell+0.20,\y);
}


\draw[red!70,line width=1pt]
    (\xstart+2*\cell,\yhigh)
    --
    (\xstart+1*\cell,\ylow+\height);

\draw[red!70,line width=1pt]
    (\xstart+3*\cell,\yhigh)
    --
    (\xstart+4*\cell,\ylow+\height);


\node[font=\large]
    at (\xstart+1.5*\cell,\ylow+0.5*\height)
    {$\mathcal J_1$};

\node[font=\large]
    at (\xstart+2.5*\cell,\ylow+0.5*\height)
    {$\mathcal J_2$};

\node[font=\large]
    at (\xstart+3.5*\cell,\ylow+0.5*\height)
    {$\mathcal J_3$};


\node[font=\large]
    at (\xstart-0.75,\ylow+0.5*\height)
    {$t$};

\node[font=\large]
    at (\xstart-0.75,\yhigh+0.5*\height)
    {$t+1$};

\node[font=\large]
    at (\xstart+2.5*\cell,\yhigh+\height+0.38)
    {$\Omega$};

\node[font=\large]
    at (\xstart+2.5*\cell,\ylow-0.40)
    {$N_{\mathbf q}(\Omega)$};

\node[font=\Large]
    at (\xstart+2.5*\cell,1.15)
    {$\mathbf q$};

\end{tikzpicture}%
}

\caption{
A diagramatic representation of an OMP.
}
\label{fig:omp_causal_neighbourhood}

\end{figure}

The protocol contains \(n\) applications of
the QCA dynamics and therefore has temporal duration \(n\). The final readout is performed at time \(t+n\) on the localized region
\(\Omega_0\). The branch \(\mathcal{J}_{\Omega_0}^{m_0}\) of the readout layer determines the local readout effect
\begin{equation}
    E_{\Omega_0}^{m_0}
    :=
    \mathcal{J}_{\Omega_0}^{m_0}
    (\mathbf{1}_{\Omega_0})
    \in
    \alg A_{\Omega_0}.
    \label{eq:local-readout-effect}
\end{equation} To compare the Schrödinger and Heisenberg descriptions, separate the final readout from the preceding part of the protocol. Define
\begin{equation}
    \mathcal{R}_{\QCA{q}}^{m_{>0}}
    :=
    \mathcal{J}_{\Omega_n}^{m_n}
    \circ
    \QCA{q}
    \circ
    \mathcal{J}_{\Omega_{n-1}}^{m_{n-1}}
    \circ\cdots\circ
    \QCA{q}
    \circ
    \mathcal{J}_{\Omega_1}^{m_1}
    \circ
    \QCA{q},
    \label{eq:omp-pre-readout-map}
\end{equation}
where $m_{>0} := (m_1,\ldots,m_n)$, The complete branch then factorizes as $\mathcal{P}_{\QCA{q}}^{m} = \mathcal{R}_{\QCA{q}}^{m_{>0}} \circ \mathcal{J}_{\Omega_0}^{m_0}$.

\smallskip 
Let \(\omega_t\) be the state of the lattice at the initial time \(t\). In the Schrödinger picture, the operations supported on
\(\Omega_n,\ldots,\Omega_1\), interleaved with the \(n\) QCA steps, transform \(\omega_t\) into the subnormalized pre-readout state
\begin{equation}
    \eta_{t+n}^{m_{>0}}
    :=
    \omega_t
    \circ
    \mathcal{R}_{\QCA{q}}^{m_{>0}}.
    \label{eq:pre-readout-state}
\end{equation} The final measurement is then performed on $\Omega_0$. The
subnormalized state associated with the complete outcome $\mathbf m=(m_0,\mathbf m_{>0})$ is therefore
\begin{equation}
    \eta^{\mathbf m}_{t+n}
    :=
    \eta^{\mathbf m_{>0}}_{t+n}
    \circ
    \mathcal J_{\Omega_0}^{m_0}
    =
    \omega_t
    \circ
    \mathcal P_{\QCA q}^{\mathbf m}.
    \label{eq:complete-subnormalized-state}
\end{equation} Its normalization gives the joint probability of the complete outcome
string:
\[
    p(\mathbf m\mid\omega_t)
    =
    \eta^{\mathbf m}_{t+n}(\mathbf 1)
    =
    \eta^{\mathbf m_{>0}}_{t+n}
    \left(
        E_{\Omega_0}^{m_0}
    \right).
\] In the Heisenberg picture, the initial state $\omega_t$ is kept fixed
and the local readout effect is propagated through the adjoint
description of the complete spacetime protocol. The resulting
effective effect on the initial time slice is
\[
    E_{\Omega_n}^{\mathbf m}
    :=
    \mathcal R_{\QCA q}^{\mathbf m_{>0}}
    \left(
        E_{\Omega_0}^{m_0}
    \right)
    =
    \mathcal P_{\QCA q}^{\mathbf m}(\mathbf 1).
\] Consequently,
\begin{equation}
    p(\mathbf m\mid\omega_t)
    =
    \eta^{\mathbf m}_{t+n}(\mathbf 1)
    =
    \eta^{\mathbf m_{>0}}_{t+n}
    \left(
        E_{\Omega_0}^{m_0}
    \right)
    =
    \omega_t
    \left(
        E_{\Omega_n}^{\mathbf m}
    \right).
    \label{eq:omp-probability}
\end{equation} Equation \eqref{eq:omp-probability} is the operational content of the
construction. Measuring the local effect $E_{\Omega_0}^{m_0}$ at time $t+n$ after applying the distributed protocol produces exactly the same statistics as evaluating the effective effect $E_{\Omega_n}^{\mathbf m}$ at time $t$ in the original state $\omega_t$. Whenever $p(\mathbf m\mid\omega_t)>0$, the normalized conditional post-measurement state associated with the complete outcome $\mathbf m$ is
\begin{equation}
    \omega^{\mathbf m}_{t+n}
    :=
    \frac{
        \eta^{\mathbf m}_{t+n}
    }{
        p(\mathbf m\mid\omega_t)
    }.
    \label{eq:conditional-state-omp}
\end{equation}

The localization of the effective effect follows directly from QCA causality. Starting from $E_{\Omega_0}^{m_0} \in \alg A_{\Omega_0}$ one has, at every step,
\[
    \QCA{q}(\alg A_{\Omega_{\ell-1}})
    \subseteq
    \alg A_{\Omega_\ell},
    \qquad
    \Omega_\ell
    =
    N_{\QCA{q}}(\Omega_{\ell-1}).
\] The subsequent instrument branch
\(\mathcal{J}_{\Omega_\ell}^{m_\ell}\) maps
\(\alg A_{\Omega_\ell}\) into itself. Induction therefore gives
\begin{equation}
    E_{\Omega_n}^{m}
    \in
    \alg A_{\Omega_n},
    \qquad
    \Omega_n
    =
    N_{\QCA{q}}^{n}(\Omega_0).
    \label{eq:effective-effect-support}
\end{equation} Accordingly, time is an explicit resource of the protocol. Each application of \(\QCA{q}\) permits the readout effect to spread by one causal neighbourhood. An OMP of duration \(n\) can therefore reconstruct only effects supported inside $N_{\QCA{q}}^{n}(\Omega_0)$. In particular, a necessary causal condition for reconstructing an effect supported in a finite region \(\Lambda\) is $\Lambda \subseteq N_{\QCA{q}}^{n}(\Omega_0)$. This condition is necessary but not sufficient: actual implementability also requires that the target effect admit the appropriate factorization through
the local instrument branches and the QCA dynamics.

\smallskip An OMP is therefore a spacetime-distributed measurement architecture. Its elementary interventions are strictly local, but they are distributed over the causally admissible regions at successive times. The QCA dynamics is the only mechanism by which
information is transferred between different layers. The extended support of the effective measurement is obtained by using both the spatial degrees of freedom and the finite time available for their causal interaction.

\subsubsection{Operational Measurement Protocols and Causality}
\label{subsec:discrete-three-region-causality}

We now examine the causal admissibility of the local operations employed in an Operational Measurement Protocol. The relevant obstruction is the discrete counterpart of the three-region scenario: an operation localized in an extended intermediate region may, in principle, couple two causally distinct components of that region and thereby create a signalling path which is not generated by the QCA dynamics. 

\smallskip 
Throughout this subsection we use the Heisenberg convention. If an observable is measured at a given time, successive applications of \(\QCA{q}\) provide its representatives at earlier times. Equivalently, if \(\omega_t\) is the state at time \(t\), then
\[
    \omega_{t+1}(A)
    =
    (\omega_t \circ \QCA{q})(A).
\] Thus \(\QCA{q}(A)\) is not the observable \(A\) evolved towards a later time: it is the representative at time \(t\) of the observable \(A\) measured at time \(t+1\).

\smallskip 
Let $t_A<t_C<t_B$ be three discrete times, and let $\Omega_A,\Omega_C,\Omega_B \Subset\mathbb Z^D$ be finite regions localized at the respective times. Let $B\in\alg A_{\Omega_B}$ be an observable measured in \(\Omega_B\) at time \(t_B\). Its Heisenberg
representative at the intermediate time \(t_C\) is
\[
    B_C
    :=
    \QCA{q}^{t_B-t_C}(B)
    \in
    \alg A_{N_{\QCA{q}}^{t_B-t_C}(\Omega_B)}.
\] In the absence of an intermediate intervention, its representative at time \(t_A\) is
\[
    B_A
    :=
    \QCA{q}^{t_B-t_A}(B)
    \in
    \alg A_{N_{\QCA{q}}^{t_B-t_A}(\Omega_B)}.
\] Accordingly, the spacetime regions
\((\Omega_A,t_A)\) and \((\Omega_B,t_B)\) are causally disjoint whenever
\begin{equation}
    \Omega_A
    \cap
    N_{\QCA{q}}^{t_B-t_A}(\Omega_B)
    =
    \varnothing.
    \label{eq:three-region-causal-disjointness}
\end{equation} Under this condition, $B_A$ is localized in a region disjoint from \(\Omega_A\). Consequently, every deterministic operation \(\mathcal C_{\Omega_A}\) localized in \(\Omega_A\) satisfies $\mathcal C_{\Omega_A}(B_A)=B_A$. It follows that, for every state \(\omega_{t_A}\),
\[
    \omega_{t_A}
    \left(
        \mathcal C_{\Omega_A}(B_A)
    \right)
    =
    \omega_{t_A}(B_A).
\] Therefore an intervention performed in \(\Omega_A\) at time \(t_A\) cannot change the statistics of \(B\) at time \(t_B\).

\smallskip 
Suppose now that a deterministic quantum operation $\mathcal J_{\Omega_C}:\alg A\to\alg A$ localized in \(\Omega_C\) is applied at time \(t_C\). The observable at time \(t_A\) induced by the complete sequence is
\begin{align*}
    B_A^{(\mathcal J_{\Omega_C})}
    & :=
    \big(\QCA{q}^{t_C-t_A} \circ \mathcal J_{\Omega_C} \circ  \QCA{q}^{t_B-t_C}\big)(B) \\[1mm]
    & =
    \QCA{q}^{t_C-t_A}
    \big(
        \mathcal J_{\Omega_C}(B_C)
    \big). \notag
\end{align*} For an initial state \(\omega_{t_A}\), the expectation value measured at the final time is therefore
\[
    \omega_{t_B}^{(\mathcal J_{\Omega_C})}(B)
    =
    \omega_{t_A}
    \left(
        B_A^{(\mathcal J_{\Omega_C})}
    \right).
\] An operation in \(\Omega_A\), performed before the intermediate operation, would instead give $(\omega_{t_A} \circ \mathcal C_{\Omega_A})(B_A^{(\mathcal J_{\Omega_C})})$. The intermediate operation produces the three-region signalling pathology if, despite Equation~\eqref{eq:three-region-causal-disjointness}, there exist \(B\in\alg A_{\Omega_B}\), a deterministic operation \(\mathcal C_{\Omega_A}\), and a state \(\omega_{t_A}\) such that
\begin{equation}
    \omega_{t_A}
    \left(
        \mathcal C_{\Omega_A}
        \left(
            B_A^{(\mathcal J_{\Omega_C})}
        \right)
    \right)
    \neq
    \omega_{t_A}
    \left(
        B_A^{(\mathcal J_{\Omega_C})}
    \right).
    \label{eq:three-region-pathology}
\end{equation} In this case, the operation performed in the causally disjoint region \(\Omega_A\) changes the statistics subsequently observed in \(\Omega_B\).

\smallskip
To identify the mechanism by which this may happen, define the two relevant components of the intermediate region:
\begin{align}
    \Omega_C^B
    &:=
    \Omega_C
    \cap
    N_{\QCA{q}}^{t_B-t_C}(\Omega_B),
    \label{eq:three-region-component-B}\\
    \Omega_C^A
    &:=
    \left\{
        x_C\in\Omega_C:
        \Omega_A
        \cap
        N_{\QCA{q}}^{t_C-t_A}(\{x_C\})
        \neq\varnothing
    \right\}.
    \label{eq:three-region-component-A}
\end{align} The set \(\Omega_C^B\) contains the sites of the intermediate region on which \(B_C\) may be supported. The set \(\Omega_C^A\) contains the sites whose observables may acquire support in \(\Omega_A\) after \(t_C-t_A\) further Heisenberg steps.

\smallskip
Causal disjointness implies $\Omega_C^A\cap\Omega_C^B=\varnothing$. Indeed, suppose that some \(x_C\) belonged to both sets. Then
\[
    x_C
    \in
    N_{\QCA{q}}^{t_B-t_C}(\Omega_B), \text{ and } \Omega_A
    \cap
    N_{\QCA{q}}^{t_C-t_A}(\{x_C\})
    \neq
    \varnothing.
\] Since neighbourhood thickenings compose according to
\[
    \big(N_{\QCA{q}}^{t_C-t_A} \circ 
        N_{\QCA{q}}^{t_B-t_C}\big)(\Omega_B)
    =
    N_{\QCA{q}}^{t_B-t_A}(\Omega_B),
\] this would imply $\Omega_A \cap N_{\QCA{q}}^{t_B-t_A}(\Omega_B) \neq \varnothing$, contradicting
Equation~\eqref{eq:three-region-causal-disjointness}. The possible obstruction is now explicit. Although the component of \(B_C\) contained in \(\Omega_C\) is localized in \(\Omega_C^B\), a general channel localized on the whole algebra \(\alg A_{\Omega_C}\) need not preserve that smaller localization region. In principle, it may satisfy 
\[
    \mathcal J_{\Omega_C}
    \left(
        \alg A_{\Omega_C^B}
    \right)
    \not\subseteq
    \alg A_{\Omega_C^B}
\] and may produce an observable having a nontrivial component in
\(\Omega_C^A\). The subsequent application of
\(\QCA{q}^{t_C-t_A}\) can then carry that component into \(\Omega_A\). The intermediate operation would thereby join the two disjoint components \(\Omega_C^B\) and \(\Omega_C^A\), creating a causal path which is not generated by the QCA dynamics alone. 

Notice that the relevant obstruction is the ability of a primitive operation on \(\alg A_{\Omega_C}\) to transfer the dependence of an observable between distinct sites of that region. It is therefore unavailable when the intermediate region $\Omega_C$ consists of a single site.

\begin{proposition}
\label{prop:single-site-three-region-causality}

Assume that \((\Omega_A,t_A)\) and \((\Omega_B,t_B)\) satisfy
Equation~\eqref{eq:three-region-causal-disjointness}. Let
\(\Omega_C=\{x_C\}\), and let \(\mathcal J_{\{x_C\}}\) be a deterministic quantum operation localized in \(\{x_C\}\). Then
\[
    B_A^{(\mathcal J_{\{x_C\}})}
    \in
    \alg A_{N_{\QCA{q}}^{t_B-t_A}(\Omega_B)}
    \qquad
    \forall B\in\alg A_{\Omega_B}.
\] Consequently, \(\mathcal J_{\{x_C\}}\) cannot generate the three-region signalling pathology.
\end{proposition}

\begin{proof}
Fix \(B\in\alg A_{\Omega_B}\), and set $S_C := N_{\QCA{q}}^{t_B-t_C}(\Omega_B)$. By QCA causality, $B_C = \QCA{q}^{t_B-t_C}(B) \in \alg A_{S_C}$. We distinguish two cases. If \(x_C\notin S_C\), then the localized operation acts trivially on \(B_C\), and hence $\mathcal J_{\{x_C\}}(B_C)=B_C$. Suppose instead that \(x_C\in S_C\). Using the canonical decomposition $\alg A_{S_C} \simeq \alg A_{\{x_C\}} \otimes \alg A_{S_C\setminus\{x_C\}}$, the restriction of the localized operation is
\[
    \mathcal J_{\{x_C\}}
    \big|_{\alg A_{S_C}}
    =
    \mathcal J_{\{x_C\}}
    \otimes
    \operatorname{id}_{S_C\setminus\{x_C\}}.
\] Therefore $\mathcal J_{\{x_C\}}(B_C) \in \alg A_{S_C}$. More generally, in both cases,
\[
    \mathcal J_{\{x_C\}}
    \left(
        \alg A_{S_C}
    \right)
    \subseteq
    \alg A_{S_C}.
\] Applying the remaining \(t_C-t_A\) QCA steps and using causality gives
\begin{align*}
    B_A^{(\mathcal J_{\{x_C\}})} & =
    \QCA{q}^{t_C-t_A}
    \left(
        \overline{\mathcal J}_{\{x_C\}}(B_C)
    \right) \notag \\[1mm]
    & \in
    \alg A_{N_{\QCA{q}}^{t_C-t_A}(S_C)}=
    \alg A_{N_{\QCA{q}}^{t_B-t_A}(\Omega_B)}.
\end{align*} By Equation~\eqref{eq:three-region-causal-disjointness}, this localization region is disjoint from \(\Omega_A\). Hence every deterministic operation \(\mathcal C_{\Omega_A}\) localized in \(\Omega_A\) satisfies
\[
    \mathcal C_{\Omega_A}
    \left(
        B_A^{(\overline{\mathcal J}_{\{x_C\}})}
    \right)
    =
    B_A^{(\overline{\mathcal J}_{\{x_C\}})}.
\] Equation~\eqref{eq:three-region-pathology} can therefore never hold. Equivalently, in terms of the intermediate components, the single site
\(x_C\) cannot belong simultaneously to the disjoint sets \(\Omega_C^B\) and \(\Omega_C^A\). A single-site operation may modify the local component already present at \(x_C\), but it cannot enlarge the support to any further site.
\end{proof}

An instrument layer is a parallel composition of strictly single-site
instruments. Its non-selective channel therefore retains the support
preservation property established above for a single-site operation.

\begin{theorem}
\label{thm:instrument-layer-support-preservation}

Let $\texttt{J}_{\Omega_C}$ be an instrument layer as in
Definition~\ref{def:instrument-layer}, and let $\overline{\mathcal J}_{\Omega_C}$ be its non-selective channel. Then, for every finite region
\(S\Subset\mathbb Z^D\),
\begin{equation}
    \overline{\mathcal J}_{\Omega_C}
    \left(
        \alg A_S
    \right)
    \subseteq
    \alg A_S.
    \label{eq:instrument-layer-support-preservation}
\end{equation} Thus the non-selective channel of an instrument layer cannot transfer the support of an observable between distinct lattice sites.
\end{theorem}

\begin{proof}

Let \(A\in\alg A_S\). Since all local algebras are finite-dimensional, it is sufficient by linearity to consider elementary tensors with respect to the single-site decomposition of \(\alg A_S\). We may therefore write
\[
    A
    =
    \Big(
        \bigotimes_{x_C\in S\cap\Omega_C} A_{x_C}
    \Big)
    \otimes
    A_{S\setminus\Omega_C}.
\] When \(A\) is embedded into \(S\cup\Omega_C\), its factor at every site \(x_C\in\Omega_C\setminus S\) is the identity \(\mathbf 1_{\{x_C\}}\). For \(x_C\in S\cap\Omega_C\), the corresponding local channel satisfies $\overline{\mathcal J}_{\{x_C\}}(A_{x_C}) \in \alg A_{\{x_C\}}$, so the local factor remains supported at \(x_C\). For
\(x_C\in\Omega_C\setminus S\), unitality gives $\overline{\mathcal J}_{\{x_C\}}( \mathbf 1_{\{x_C\}}) = \mathbf 1_{\{x_C\}}$. Finally, the layer acts identically on the tensor factors associated with
\(S\setminus\Omega_C\). Consequently,
\[
    \overline{\mathcal J}_{\Omega_C}(A)
    =
    \Big(
        \bigotimes_{x_C\in S\cap\Omega_C}
        \overline{\mathcal J}_{\{x_C\}}(A_{x_C})
    \Big)
    \otimes
    A_{S\setminus\Omega_C}
    \in
    \alg A_S.
\] Since elementary tensors span \(\alg A_S\), the result follows for every \(A\in\alg A_S\).
\end{proof}

We can now apply the support-preservation theorem directly to the
three-region scenario.

\begin{theorem}
\label{thm:instrument-layer-three-region-causality}

Assume that \((\Omega_A,t_A)\) and \((\Omega_B,t_B)\) satisfy
Equation~\eqref{eq:three-region-causal-disjointness}. Let
\(\texttt{J}_{\Omega_C}\) be an instrument layer applied at time \(t_C\), and let \(\mathcal J_{\Omega_C}\) be its non-selective channel. Then, for every \(B\in\alg A_{\Omega_B}\),
\begin{equation}
    B_A^{(\mathcal J_{\Omega_C})}
    \in
    \alg A_{N_{\QCA{q}}^{t_B-t_A}(\Omega_B)}.
    \label{eq:instrument-layer-induced-support}
\end{equation} Consequently, the instrument layer cannot generate the three-region signalling pathology.
\end{theorem}

\begin{proof}

Let $B_C = \QCA{q}^{t_B-t_C}(B)$. By QCA causality,
\[
    B_C
    \in
    \alg A_{N_{\QCA{q}}^{t_B-t_C}(\Omega_B)}.
\]
Applying Theorem~\ref{thm:instrument-layer-support-preservation} with $S= N_{\QCA{q}}^{t_B-t_C}(\Omega_B)$ gives $\mathcal J_{\Omega_C}(B_C) \in \alg A_{N_{\QCA{q}}^{t_B-t_C}(\Omega_B)}$. The layer may therefore modify \(B_C\) independently at the sites already
contained in its localization region, but it cannot extend its support to new sites. In particular, it cannot transfer the component of \(B_C\) contained in \(\Omega_C^B\) to the disjoint component \(\Omega_C^A\). Applying the remaining QCA evolution gives
\begin{align*}
    B_A^{(\mathcal J_{\Omega_C})}
    &=
    \QCA{q}^{t_C-t_A}
    \left(
        \mathcal J_{\Omega_C}(B_C)
    \right) \\
    & \in \alg A_{N_{\QCA{q}}}^{t_C-t_A}
        \left(
            N_{\QCA{q}}^{t_B-t_C}(\Omega_B)
        \right) =
    \alg A_{N_{\QCA{q}}^{t_B-t_A}(\Omega_B)}.
\end{align*} This proves Equation~\eqref{eq:instrument-layer-induced-support}. Since the latter region is disjoint from \(\Omega_A\), every deterministic operation \(\mathcal C_{\Omega_A}\) satisfies
\[
    \mathcal C_{\Omega_A}
    \left(
        B_A^{(\mathcal J_{\Omega_C})}
    \right)
    =
    B_A^{(\mathcal J_{\Omega_C})}.
\] It follows that
\[
    (\omega_{t_A} \circ 
        \mathcal C_{\Omega_A})
    \left(
            B_A^{(\mathcal J_{\Omega_C})}
        \right)
    =
    \omega_{t_A}
    \left(
        B_A^{(\mathcal J_{\Omega_C})}
    \right)
\] for every state \(\omega_{t_A}\). Hence the statistics observed in
\(\Omega_B\) cannot depend on a deterministic intervention performed in
\(\Omega_A\).
\end{proof}

The preceding statement concerns the non-selective channel associated with the instrument layer. This qualification is essential. A selective branch \(\mathcal J_{\Omega_C}^{\mathbf m_C}\) is generally subunital rather than unital. Consequently, for a site \(x_C\notin S\), one generally has
\[
    \mathcal J_{\{x_C\}}^{m_{x_C}}
    \left(
        \mathbf 1_{\{x_C\}}
    \right)
    =
    E_{\{x_C\}}^{m_{x_C}}
    \neq
    \mathbf 1_{\{x_C\}}.
\] A selective branch may therefore produce an effect on sites that were not contained in the support of the original observable. This represents conditioning on the recorded outcomes and does not, by itself, constitute operational signalling. The outcome must be communicated and used for post-selection in order for the corresponding conditional dependence to be accessible. No-signalling is therefore a property of the non-selective channel.

\smallskip

An OMP consists of instrument layers interleaved with applications of the QCA dynamics. Each non-selective instrument layer preserves the support of an observable, whereas every QCA step enlarges that support only according to the neighbourhood scheme \(N_{\QCA{q}}\). The complete non-selective protocol consequently inherits the causal propagation bound of the QCA.

\begin{theorem}
\label{thm:omp-causal-admissibility}

Let \(\texttt{P}_{\QCA{q}}\) be an OMP as in
Definition~\ref{def:omp}, and let \(\overline{\mathcal P}_{\QCA{q}}\) denote its non-selective channel. Then, for every finite region \(S\Subset\mathbb Z^D\),
\begin{equation}
    \overline{\mathcal P}_{\QCA{q}}
    \left(
        \alg A_S
    \right)
    \subseteq
    \alg A_{N_{\QCA{q}}^{n}(S)}.
    \label{eq:omp-nonselective-causal-support}
\end{equation} Consequently, the non-selective channel of an OMP cannot generate a causal connection outside the cone generated by the QCA dynamics and cannot produce the three-region signalling pathology.
\end{theorem}

\begin{proof}

Let the non-selective instrument layers of the protocol be $\mathcal J_{\Omega_0}, \mathcal J_{\Omega_1}, \ldots, \mathcal J_{\Omega_n}$, so that
\begin{equation*}
    \overline{\mathcal P}_{\QCA{q}}
    =
    \overline{\mathcal J}_{\Omega_n}
    \circ
    \QCA{q}
    \circ
    \overline{\mathcal J}_{\Omega_{n}}
    \circ\cdots\circ
    \QCA{q}
    \circ
    \overline{\mathcal J}_{\Omega_0}.
\end{equation*} Fix \(A\in\alg A_S\). By
Theorem~\ref{thm:instrument-layer-support-preservation}, $ \overline{\mathcal J}_{\Omega_0}(A) \in \alg A_S$. The first QCA step therefore gives
\[
    \QCA{q}
    \left(
        \overline{\mathcal J}_{\Omega_0}(A)
    \right)
    \in
    \alg A_{N_{\QCA{q}}(S)}.
\] The next instrument layer preserve this enlarged support:
\[
    \overline{\mathcal J}_{\Omega_1}
    \circ
    \QCA{q}
    \circ
    \overline{\mathcal J}_{\Omega_0}(A)
    \in
    \alg A_{N_{\QCA{q}}(S)}.
\] Suppose inductively that, after \(k\) QCA steps and the subsequent
instrument layer, the resulting observable belongs to $\alg A_{N_{\QCA{q}}^{k}(S)}$. The next application of \(\QCA{q}\) maps it into $\alg A_{N_{\QCA{q}}^{k+1}(S)}$ and the following instrument layer preserves this support. By induction, after \(n\) QCA steps,
\[
    \overline{\mathcal P}_{\QCA{q}}(A)
    \in
    \alg A_{N_{\QCA{q}}^{n}(S)}.
\]
This proves Equation~\eqref{eq:omp-nonselective-causal-support}. 

Now let \(B\in\alg A_{\Omega_B}\) be a final observable and suppose that an OMP of temporal depth \(n=t_B-t_A\) is applied between \(t_A\) and \(t_B\). Its representative at the initial time is
\[
    B_A^{(\texttt{P}_{\QCA{q}})}
    :=
    \overline{\mathcal P}_{\QCA{q}}(B)
    \in
    \alg A_{N_{\QCA{q}}^{t_B-t_A}(\Omega_B)}.
\] If $\Omega_A \cap N_{\QCA{q}}^{t_B-t_A}(\Omega_B) = \varnothing$, then every deterministic operation \(\mathcal C_{\Omega_A}\) localized in \(\Omega_A\) leaves this induced observable invariant. Consequently,
\[
    (\omega_{t_A} \circ 
        \mathcal C_{\Omega_A})  
        \left(
            B_A^{(\texttt{P}_{\QCA{q}})}
        \right)
    =
    \omega_{t_A}
    \left(
        B_A^{(\texttt{P}_{\QCA{q}})}
    \right)
\] for every initial state \(\omega_{t_A}\). The OMP therefore cannot produce signalling between causally disjoint spacetime regions.
\end{proof}

Equation~\eqref{eq:omp-nonselective-causal-support} shows that every transfer of support occurring in a non-selective OMP is generated by an explicit application of the QCA dynamics. Instrument layers can modify the observable independently at each site, but they cannot transfer it between distinct sites or join causally disjoint components of an intermediate region. This causal admissibility does not imply that the effective effects reconstructed by the selective branches of an OMP must factorize spatially. Although each instrument layer is assembled from strictly single-site operations, the intervening QCA dynamics may couple observables initially localized at different sites. The Heisenberg representative of a final local readout effect can therefore be an extended and non-factorizing effect supported in $N_{\QCA{q}}^{n}(\Omega_0)$. Its non-factorization is generated through the allowed QCA evolution over \(n\) discrete time steps; it is not produced by a primitive operation coupling causally disconnected components at a single time. Thus an OMP may reconstruct the statistics of spatially extended and quantum-correlated effects while remaining compatible with the causal structure determined by \(N_{\QCA{q}}\).

\section{The Bell Operational Measurement Protocol}
\label{sec:bell-omp}

We now instantiate the general construction in the simplest setting in which an Operational Measurement Protocol reconstructs a genuinely entangled measurement, that is, a one-dimensional nearest-neighbour
qubit QCA with trivial index. We use the notation $\QCA{q}$ for the QCA automorphism, $N_\QCA{q}=\{-1,0,+1\}$ for its neighbourhood scheme, and $\operatorname{CPZ}(\varphi)$ for the controlled-phase gate appearing in the normal form of Theorem \ref{thm:qubit-qca-classification}.

\smallskip 
Let $\Omega_0:=\{x\}$ and, for $r\geq 0$, let
$\Omega_r:=N_\QCA{q}^r(\Omega_0)$. Thus,
\[
\begin{aligned}
    \Omega_1 & =\{x-1,x,x+1\}, \\[1mm]
    \Omega_2 & =\{x-2,x-1,x,x+1,x+2\}.
\end{aligned} 
\] We use the notation $\{y\}_r$ for the copy of the site $y$ belonging to the $r$-th instrument layer. This layer label distinguishes spacetime locations in the protocol; the corresponding local observable algebra is still $\mathsf A_{\{y\}}$.

The protocol is the two-step OMP
\[
    \texttt{P}_\QCA{q}
    =
    \texttt{J}_{\Omega_2} \circ \QCA{q} \circ \texttt{J}_{\Omega_1} \circ \QCA{q} \circ \texttt{J}_{\Omega_0},
\] with complete branch label $m=(n,j,i,\ell)\in\mathbb{Z}_2^4$. For a single Kraus operator $K$, we write $\operatorname{Ad}[K](A):=K^\dagger A K$. The readout layer is the computational-basis PVM on $\Omega_0$:
\[
    \texttt{J}_{\Omega_0}
    =
    \bigl\{\mathcal J_{\Omega_0}^{(n)}\bigr\}_{n\in\mathbb{Z}_2},
    \quad
    \mathcal J_{\Omega_0}^{(n)}
    =
    \operatorname{Ad}\left[\proj{n}_{\{x\}}\right].
\] The first instrument layer is the tensor product of the following
single-site instruments:
\[
\begin{aligned}
    \mathcal J_{\{x-1\}_1}^{(j)}
    &:=
    \operatorname{Ad}\left[\proj{j}_{\{x-1\}}\right],
    &&j\in\mathbb{Z}_2,
    \\[1mm]
    \mathcal J_{\{x\}_1}
    &:=
    \operatorname{Ad}\left[H_{\{x\}}\right],
    \\[1mm]
    \mathcal J_{\{x+1\}_1}^{(i)}
    &:=
    \operatorname{Ad}\left[H_{\{x+1\}}
    \proj{i}_{\{x+1\}}\right],
    &&i\in\mathbb{Z}_2,
\end{aligned}
\] where $H$ is the Hadamard gate,
\[
H = \frac{1}{\sqrt{2}}\begin{pmatrix}
    1 & 1 \\ 1 & -1
\end{pmatrix}.
\] Hence $\mathcal J_{\Omega_1}^{(j,i)} := \mathcal J_{\{x-1\}_1}^{(j)} \otimes \mathcal J_{\{x\}_1} \otimes \mathcal J_{\{x+1\}_1}^{(i)}$. The second layer contains the identity instrument on the first three sites, a deterministic Hadamard operation on $\{x+1\}_2$, and a computational-basis PVM on $\{x+2\}_2$:
\[
\begin{aligned}
    \mathcal J_{\{x+1\}_2}&:=
    \operatorname{Ad}\left[H_{\{x+1\}}\right],
    \\[1mm]
    \mathcal J_{\{x+2\}_2}^{(\ell)}&:=
    \operatorname{Ad}\left[\proj{\ell}_{\{x+2\}}\right],
    &\ell\in\mathbb{Z}_2.
\end{aligned}
\] Thus $\mathcal J_{\Omega_2}^{(\ell)} := \mathcal J_{\{x+1\}_2} \otimes \mathcal J_{\{x+2\}_2}^{(\ell)}$. With the Heisenberg ordering fixed by the definition of an OMP, the branch operation is and its effective effect are
\[
    \mathcal P_\QCA{q}^{(n,j,i,\ell)}
    :=
    \mathcal J_{\Omega_2}^{(\ell)}\circ \QCA{q}\circ
    \mathcal J_{\Omega_1}^{(j,i)}\circ \QCA{q}\circ
    \mathcal J_{\Omega_0}^{(n)},
\] while its effective effect is
\[
    E_{\Omega_2}^{(n,j,i,\ell)}
    :=
    \mathcal P_\QCA{q}^{(n,j,i,\ell)}(\mathbf 1)
\] For every initial state $\omega$ on the quasi-local algebra,
$p(n,j,i,\ell\mid\omega)=\omega(E_{\Omega_2}^{(n,j,i,\ell)})$.

\smallskip 

At $\varphi=\pi$, the controlled-phase interaction is the controlled-Z
gate. The explicit calculation in Appendix~\ref{sec:supplementary-materials}
gives $E_{\Omega_2}^{(n,j,i,\ell)}$ equal to
\[
    \proj{j}_{\{x-1\}}
    \otimes\proj{\beta^{(n\oplus j,\,i\oplus\ell)}}_{\{x,x+1\}}
    \otimes\proj{\ell}_{\{x+2\}}.
\]
Here $\oplus$ denotes addition modulo $2$, and the Bell vectors are indexed
as in the appendix so that the displayed circuit identity holds. Finally, defining the coarse-graining map
\[
    F : \mathbb{Z}_2^4 \to \mathbb{Z}_2^2, \qquad F(n,j,i,\ell):=(n\oplus j,i\oplus\ell),
\] and summing the fine-grained
effects over $F^{-1}(a,b)$ gives
\[
    E_{\Omega_2}^{(a,b)}
    =
    \mathbf 1_{\{x-2,x-1\}}
    \otimes\proj{\beta^{(a,b)}}_{\{x,x+1\}}
    \otimes\mathbf 1_{\{x+2\}}.
\] After removing the identity factors on the record sites, this is the Bell
PVM on the central pair.

\smallskip

We now consider a general interaction phase $\varphi\in(-\pi,\pi]$. The explicit calculation is given in Appendix~\ref{sec:supplementary-materials}. The outer outcomes $j$ and $\ell$ act as controls on the two central sites. Conditioning one input of a controlled-phase gate on a computational-basis outcome
$r\in\mathbb Z_2$ induces, up to a global phase, the single-qubit rotation $R_Z(\theta)^r$,
\[
    R_Z(\theta):=e^{-i\theta Z/2},
\] on the other input. The effective unitary acting on the pair of sites $\{x,x+1\}$
is therefore
\[
    U^{(\ell,j)}(\varphi)_{\{x,x+1\}}
    :=
    \operatorname{CPZ}(\varphi)_{\{x,x+1\}}
    R_Z(\varphi)^j_{\{x\}}
    R_Z(\varphi)^\ell_{\{x+1\}} .
\]
All three factors are diagonal in the computational basis and hence
commute. To isolate the deviation from the Bell point, define
\[
    \delta:=\varphi-\pi, \qquad a:=n\oplus j,
    \qquad
    b:=i\oplus\ell.
\] For a complete outcome $(n,j,i,\ell)$, we define the corresponding state by
\[
    \left|\tau_{\varphi}^{(a,b;j,\ell)}\right\rangle
    :=
    H_{\{x+1\}}
    U^{(\ell,j)}(\delta)_{\{x,x+1\}}
    H_{\{x+1\}}
    \left|\beta^{(a,b)}\right\rangle .
\] Equivalently, using $R_X(\theta):= H R_Z(\theta)H = e^{-i\theta X/2}$ and
\[
    \operatorname{CPX}(\theta)_{\{x,x+1\}}
    :=
    H_{\{x+1\}}
    \operatorname{CPZ}(\theta)_{\{x,x+1\}}
    H_{\{x+1\}},
\] the same state can be written as
\[
    \left|\tau_{\varphi}^{(a,b;j,\ell)}\right\rangle
    =
    \operatorname{CPX}(\delta)_{\{x,x+1\}}
    R_Z(\delta)^j_{\{x\}}
    R_X(\delta)^\ell_{\{x+1\}}
    \left|\beta^{(a,b)}\right\rangle .
\] The corresponding fine-grained effective effect $E_{\Omega_2}^{(n,j,i,\ell)}(\varphi)$ is a projection onto the sate
\begin{equation}
\begin{aligned}
    \ket{j}_{\{x-1\}}
    \otimes
        \ket{\tau_{\varphi}^{(n\oplus j,i\oplus\ell;j,\ell)}}_{\{x,x+1\}}
    \otimes
    \ket{\ell}_{\{x+2\}} .
\end{aligned}
\label{eq:general-fine-grained-effect}
\end{equation} The dependence on the outer outcomes $(j,\ell)$ is essential away
from the Bell point. For each fixed pair $(j,\ell)$, the four states
\[
    \left\{
        \left|\tau_{\varphi}^{(a,b;j,\ell)}\right\rangle
        :
        (a,b)\in\mathbb Z_2^2
    \right\} \subset \mathbb{C}^2 \otimes \mathbb{C}^2
\] form an orthonormal basis of the central two-qubit Hilbert space,
since they are obtained from the Bell basis by the same unitary
transformation. However, for general $\varphi$ this basis itself
depends on $(j,\ell)$. Thus the central state cannot, in general, be
labelled by $(a,b)$ alone, and the coarse-graining that eliminates
the outer outcomes at the Bell point does not produce a single
$(a,b)$-labelled projective measurement away from
$\varphi=\pi$. At the Bell point the additional dependence disappears. Indeed,
for $\varphi=\pi$ one has $\delta=0$ and $U^{(\ell,j)}(0)=\mathbf 1$, so that
\[
    \left|\tau_{\pi}^{(a,b;j,\ell)}\right\rangle
    =
    \left|\beta^{(a,b)}\right\rangle
\] independently of $(j,\ell)$. The coarse-graining over the outer
records therefore reduces exactly to the Bell PVM derived above.

\smallskip 
For arbitrary $\varphi$, the protocol thus generates a family of
pure two-qubit effects continuously connected to the Bell basis,
whose structure is controlled by the interaction phase and by the
recorded outer outcomes. These states provide a natural setting in
which to quantify the quantum resources generated by the QCA
dynamics. In the following subsections we analyze two such resources
relevant to quantum information processing: bipartite entanglement,
quantified by the entropy of entanglement, and stabilizer
non-classicality, quantified by the order-two stabilizer R\'enyi
entropy \cite{LeoneOlivieroHamma2022}. This allows us to distinguish the maximally entangled Bell point from interaction phases at which the induced measurement
develops non-stabilizer character.

\subsection{Entanglement generated by the Bell OMP}

The central effect associated with each fine-grained outcome is
a rank-one projector onto a pure two-qubit state. We quantify
the entanglement of this state by the von Neumann entropy of
either one-qubit marginal, a standard measure of bipartite
entanglement for pure states \cite{Bennett1996}. Because local
unitaries preserve the Schmidt coefficients, the final local
Hadamard in the protocol does not affect this quantity.

The calculation in Appendix~\ref{sec:supplementary-materials}
gives the same marginal spectrum for every branch:
\[
    \lambda_\pm(\varphi)
    =
    \frac{1\pm|\cos(\varphi/2)|}{2}.
\] Hence the entanglement entropy is
\[
    S(\varphi)
    =
    H_2\!\left(
        \frac{1+|\cos(\varphi/2)|}{2}
    \right),
\]
where $H_2$ is the binary Shannon entropy and \(0\log_2 0:=0\). In particular, \(S(0)=0\) and \(S(\pi)=1\): the central rays are product states
at the trivial interaction and maximally entangled states at the Bell point. Moreover, for each fixed \(\varphi\), the local inputs selected by the protocol maximize the entanglement that the controlled-phase interaction can generate from product inputs. Both the spectrum and this optimization are derived in the Supplementary Material.

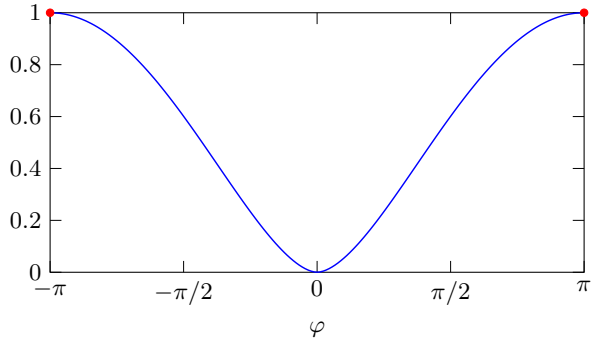
\begin{figure}[tb]
\centering
\begin{tikzpicture}
\begin{axis}[
  width=\linewidth,
  height=0.58\linewidth,
  xlabel={$\varphi$},
  ylabel={},
  xmin=-3.1415926535, xmax=3.1415926535,
  ymin=0, ymax=1,
  trig format=rad,
  samples=600,
  domain=-3.1415926535:3.1415926535,
  xtick={-3.1415926535,-1.5707963268,0,
          1.5707963268,3.1415926535},
  xticklabels={$-\pi$,$-\pi/2$,0,$\pi/2$,$\pi$},
  grid=none,
  tick style={black, line width=.2pt},
]
\addplot[blue, line width=0.55pt]
  {-((1+abs(cos(x/2)))/2)
     *log2(max((1+abs(cos(x/2)))/2,0.000000001))
   -((1-abs(cos(x/2)))/2)
     *log2(max((1-abs(cos(x/2)))/2,0.000000001))};
\addplot[only marks, mark=*, mark size=1.4pt, red]
  coordinates {(-3.1415926535,1) (3.1415926535,1)};
\end{axis}
\end{tikzpicture}
\caption{Entanglement entropy of the central branch states
as a function of the interaction phase. The value is
independent of the measurement outcomes.}
\label{fig:example_entanglement}
\end{figure}

\subsection{Magic generated by the Bell OMP}

Stabilizer quantum mechanics is the standard efficiently simulable fragment of finite-dimensional quantum theory. It is generated by Pauli strings, stabilizer states, Clifford unitaries, and Pauli measurements. Circuits built only from these ingredients admit efficient classical simulation by the Gottesman--Knill theorem \cite{Gottesman1998Heisenberg,GottesmanThesis,AaronsonGottesman}. Universal quantum computation requires leaving this sector, either by preparing non-stabilizer states or by applying non-Clifford gates. The resource responsible for this departure is usually called \emph{magic}.

\smallskip To characterize
magic, we use the stabilizer Rényi
entropy (SRE) \cite{LeoneOlivieroHamma2022}. Let \(\mathbb{P}_N\) denote a choice of representatives of the \(N\)-qubit Pauli group modulo phases. For a pure
\(N\)-qubit state \(\ket{\psi}\), its order-two SRE is
\[
    M_2[\ket{\psi}]
    =
    N-\log_2\left(
        \sum_{P\in\mathbb{P}_N}
        |\langle\psi|P|\psi\rangle|^4
    \right).
\] It vanishes on pure stabilizer states and is invariant
under Clifford unitaries.

The Supplementary Material establishes the Clifford
equivalence between the state used to evaluate \(M_2\)
and the state onto which the corresponding central effect
projects. The result depends on the outer outcomes
\((j,\ell)\), while the dependence on \(n\) and \(i\)
drops out because changing either label applies a local
Pauli. The three distinct branch profiles are shown in
Figure ~\ref{fig:example_magic}; their explicit expressions
and derivation are given in Appendix~\ref{sec:supplementary-materials}.

At \(\varphi=0\), the interaction is trivial and all the
branch states are stabilizer states. At \(\varphi=\pi\),
the protocol produces Bell states, which are also
stabilizer states. Accordingly,
\[
    M_2^{(j,\ell)}(0)
    =
    M_2^{(j,\ell)}(\pi)
    =
    0.
\] In every branch class, the maximum along this family is
reached at \(\varphi=\pm\pi/2\), where
\[
    M_2^{(j,\ell)}
    \!\left(\pm\frac{\pi}{2}\right)
    =
    \log_2\frac{16}{7}.
\] The phase that maximizes the entanglement therefore
differs from the phases that maximize the SRE: the Bell
measurement has maximal entanglement and zero SRE,
whereas the largest SRE occurs at intermediate
interaction phases.

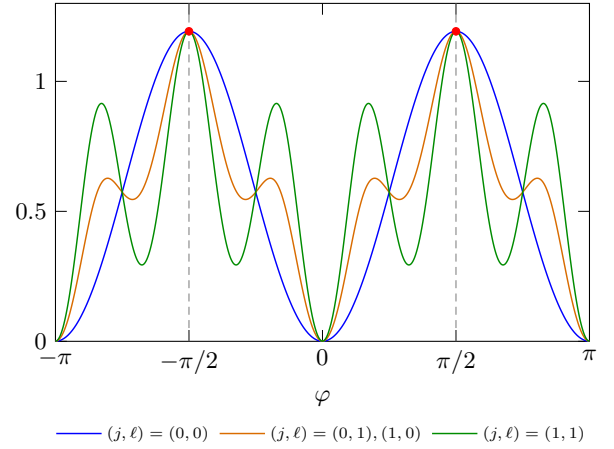
\begin{figure}[tb]
\centering
\begin{tikzpicture}
\begin{axis}[
  width=\linewidth,
  height=0.70\linewidth,
  xlabel={$\varphi$},
  ylabel={},
  xmin=-3.1415926535, xmax=3.1415926535,
  ymin=0, ymax=1.3,
  trig format=rad,
  samples=700,
  domain=-3.1415926535:3.1415926535,
  xtick={-3.1415926535,-1.5707963268,0,
          1.5707963268,3.1415926535},
  xticklabels={$-\pi$,$-\pi/2$,0,$\pi/2$,$\pi$},
  grid=none,
  tick style={black, line width=.2pt},
  legend columns=3,
  legend style={
    at={(0.5,-0.22)},
    anchor=north,
    draw=none,
    fill=none,
    font=\scriptsize,
    nodes={scale=0.85, transform shape},
    /tikz/every even column/.append style={column sep=2pt},
  },
]
\addplot[blue, line width=0.55pt]
  {2-log2(1+(3/4)*(1+cos(x)^2)^2)};
\addlegendentry{$(j,\ell)=(0,0)$}

\addplot[orange!85!black, line width=0.55pt]
  {2-log2((1/4)*
    (7+22*cos(x)^2-29*cos(x)^4
       -16*cos(x)^6+32*cos(x)^8))};
\addlegendentry{$(j,\ell)=(0,1),(1,0)$}

\addplot[green!55!black, line width=0.55pt]
  {2-log2(7/4+(19/2)*cos(x)^2+(3/4)*cos(x)^4
    -104*cos(x)^6+224*cos(x)^8
    -192*cos(x)^10+64*cos(x)^12)};
\addlegendentry{$(j,\ell)=(1,1)$}

\addplot[densely dashed, gray]
  coordinates {(-1.5707963268,0) (-1.5707963268,1.3)};
\addplot[densely dashed, gray]
  coordinates {(1.5707963268,0) (1.5707963268,1.3)};
\addplot[only marks, mark=*, mark size=1.4pt, red]
  coordinates {
    (-1.5707963268,{log2(16/7)})
    (1.5707963268,{log2(16/7)})
  };
\end{axis}
\end{tikzpicture}
\caption{Order-two stabilizer Rényi entropy of the central
branch states. The curves distinguish the three classes
of outer outcomes; the inner outcomes \(n\) and \(i\)
do not change the SRE.}
\label{fig:example_magic}
\end{figure}

\section{Discussion}

We have developed an operator-algebraic framework in which spatially
extended quantum measurements are implemented through finite sequences of
strictly local interventions and causal QCA dynamics. The central
distinction underlying the construction is between the spacetime
localization of the physical operations and the spatial support of the
effective measurement reconstructed on a reference time slice. These two
notions need not coincide: an Operational Measurement Protocol may involve
only single-site instruments at each time, while its Heisenberg pullback
defines a non-factorizing POVM supported on an extended region of the
initial lattice.

The causal structure of the QCA constrains this separation sharply. An
instrument layer acts independently on single-site tensor factors and
therefore cannot transfer the support of an observable between distinct
sites or couple causally disconnected portions of an intermediate region.
All spatial propagation of support occurring in a non-selective OMP is
instead generated by explicit applications of the QCA automorphism. As a
consequence, an order-\(n\) protocol rooted at \(\Omega_0\) can induce
effects only inside the causal neighbourhood $N_{\QCA{q}}^{\,n}(\Omega_0)$. This provides the discrete counterpart of the locality requirement exposed
by the three-region scenario: extended measurement effects are admissible
when their non-locality is generated dynamically through the available
causal propagation, rather than being introduced by a primitive operation
acting simultaneously across causally disconnected components.

The causal support bound, however, does not by itself characterize
implementability. The inclusion of the target support in
\(N_{\QCA{q}}^{\,n}(\Omega_0)\) is only a necessary condition. Whether a
given POVM can actually be synthesized also depends on whether its effects
admit a factorization through the sequence of local instrument branches
and QCA automorphisms allowed by the protocol. This separates two distinct
constraints on measurement synthesis: a geometric constraint, determined
by causal propagation, and an algebraic constraint, determined by the
structure of the dynamics and of the admissible local operations. A full
characterization of their interplay would amount to an implementability
theory for measurements associated with a fixed QCA.

The Bell protocol provides an explicit instance in which these two
requirements can be satisfied non-trivially. For the one-dimensional
nearest-neighbour qubit family considered here, the controlled-phase
interaction mediates the causal synthesis of a two-site entangled
measurement from single-site instruments. At the maximally entangling
point \(\varphi=\pi\), coarse-graining the local outcomes yields exactly
the Bell-basis projective measurement. Away from this point, the same
architecture produces a continuous family of deformed Bell measurements.
Their resource content reveals that different notions of
non-classicality need not be optimized by the same dynamics: the Bell
point maximizes the entanglement of the induced measurement while having
vanishing stabilizer Rényi non-classicality, whereas the latter reaches
its largest value within the family at intermediate interaction phases.
The automaton therefore controls not only the causal range of the
measurement, but also the type and amount of quantum resources carried by
the induced effects.

These results suggest viewing QCA dynamics itself as a resource for
measurement synthesis. Protocol duration determines how far an initially
local readout can propagate within the causal structure, whereas the
specific automorphism determines which transformations are available
inside that region. Time and interaction structure thus play logically
distinct roles: increasing the protocol depth enlarges the causally
available region, but does not by itself guarantee that every measurement
supported there becomes implementable. Determining the minimal protocol
depth required to synthesize a given measurement, when synthesis is
possible at all, is therefore a natural quantitative problem. For instance,
the Bell OPM introduced here, that provides entangled POVMs in the third step, proceeds from completely local POVMs in the first step and throug POVMs with an LOCC intepretation in the second step.

Several extensions follow from this perspective. A first problem is to
characterize, for a fixed QCA \(\QCA{q}\) and protocol order \(n\), the
set of POVMs realizable by OMPs and to identify algebraic criteria for
exact implementability. A second is to determine how this set depends on
structural invariants of the automaton, including the one-dimensional QCA
index and its higher-dimensional analogues. The present analysis is
restricted to index-one qubit automata and single-site instrument layers;
allowing non-trivial information flow, higher-dimensional local systems,
or fermionic QCAs would test which parts of the construction depend on
these restrictions and which are genuinely structural. More generally,
one may ask how the required temporal depth scales with the spatial extent
and resource content of the target measurement. 

Finally, the framework isolates a feature shared with relativistic
measurement theory while avoiding the analytic complications of
continuum AQFT: measurement-induced non-locality can be separated from
non-locality of the underlying physical intervention. QCAs therefore
provide a controlled setting in which causal admissibility and measurement
implementability can be studied independently and then related. Extending
this relation beyond discrete type-I systems---for example through
connections with detector-based measurement schemes and algebraic quantum
field theory---may clarify which aspects of the local measurement problem
are consequences of causal structure itself and which depend on the
specific algebraic properties of relativistic quantum fields.

\bibliographystyle{apsrev4-2}
\bibliography{bibliography}

\onecolumngrid
\appendix
\section*{Supplementary Material}
\label{sec:supplementary-materials}

In this Supplementary Material we give the explicit derivation of the effective POVM realized by the Bell Operational Measurement Protocol (Bell OMP), together with the closed-form expressions for the entanglement entropy and stabilizer Rényi entropy (SRE) of the two-qubit branches generated by the protocol.

\smallskip 

We use the notation of the main text. The spatial lattice is the one-dimensional qubit lattice \(\mathbb Z\), with local algebras
\[
    \mathsf A_{\{x\}}\simeq \mathsf{Mat}_2(\mathbb C),
    \qquad
    \mathsf A_{\Omega}
    =
    \bigotimes_{y\in\Omega}\mathsf A_{\{y\}}
\] for every finite region \(\Omega\Subset\mathbb Z\). In the notation $\{y\}_r$, the first entry denotes the spatial site and the second entry the instrument layer. Thus $\{y\}_r$ is a layer-labelled copy of the site $y$; the associated local algebra remains $\mathsf A_{\{y\}}$. For a single Kraus operator \(K\), we write $\operatorname{Ad}[K](A):=K^\dagger A K$ for the corresponding Heisenberg operation.

We consider a QCA $\QCA{q} \in \operatorname{Aut}(\mathsf{A})$ with a nearest-neighbour neighbourhood scheme $N_{\QCA{q}}=\{-1,0,+1\}$ with trivial index \cite{GrossNesmeVogtsWerner2012}. We fix the representative of the QCA normal form used in the
protocol. By the characterization Theorem \ref{thm:qubit-qca-classification},
\(\QCA{q}=\QCA{v}\circ\QCA{c}_\varphi\circ\QCA{u}\), where \(\QCA{u}\) and \(\QCA{v}\) are translation-invariant on-site unitary automorphisms. The interaction
\(\QCA{c}_\varphi\) can be expressed using the controlled-\(Z\)-rotation gate
\[
    \operatorname{CR}_Z(\varphi)
    :=
    \sum_{a\in\mathbb Z_2}
    \proj{a}\otimes e^{-ia\varphi Z/2}.
\] Equivalently, we may use the controlled-phase gate
\[
    \operatorname{CPZ}(\varphi)
    :=
    \operatorname{diag}(1,1,1,e^{i\varphi})
    =
    (D_\varphi\otimes\mathbf 1)
    \operatorname{CR}_Z(\varphi),
    \qquad
    D_\varphi:=\proj{0}+e^{i\varphi/2}\proj{1}.
\] Replacing \(\operatorname{CR}_Z(\varphi)\) by
\(\operatorname{CPZ}(\varphi)\) on every bond changes the
automaton only by on-site unitary automorphisms, which can
be included in \(\QCA{u}\) and \(\QCA{v}\). In the controlled-phase representative, the local update of a single-site observable is
\[
    \QCA{c}_\varphi(X_{\{y\}})
    =
    A_\varphi(y)^\dagger
    X_{\{y\}}
    A_\varphi(y),
    \qquad
    A_\varphi(y)
    :=
    \operatorname{CPZ}(\varphi)_{\{y-1,y\}}
    \operatorname{CPZ}(\varphi)_{\{y,y+1\}}.
\] In the characterization of the protocol, the on-site
automorphisms need not be written as separate evolution
layers: they can be composed with the single-site
instruments. For example, with the Heisenberg ordering
used below,
\[
    \texttt{J}_{\Omega_2}
      \circ\QCA{q}\circ
      \texttt{J}_{\Omega_1}
      \circ\QCA{q}\circ
      \texttt{J}_{\Omega_0} = (\texttt{J}_{\Omega_2}\circ\QCA{v})
      \circ\QCA{c}_\varphi \circ (\QCA{u}\circ\texttt{J}_{\Omega_1}\circ\QCA{v})
      \circ\QCA{c}_\varphi\circ
      (\QCA{u}\circ\texttt{J}_{\Omega_0}).
\] These compositions are again local instruments on the
same sites, with the same outcome labels. We therefore
give the explicit protocol in the representative
\(\QCA{q}=\QCA{c}_\varphi\); for a general automaton in
the stated normal form, the corresponding instruments
include the on-site unitary factors as above. The explicit Kraus decomposition of the protocol also involves the Hadamard gate
\[
    H = \frac{1}{\sqrt{2}} \begin{pmatrix} 1 & 1 \\ 1 & -1 \end{pmatrix}, \qquad \qquad \begin{cases}
        H\ket{0}=\ket{+}, \\[1mm]
        H\ket{1}=\ket{-}
    \end{cases}
    \qquad \ket{\pm}
    :=
    \frac{\ket{0}\pm\ket{1}}{\sqrt2}.
\]

\subsection{Detailed Construction of the Bell Operational Measurement Protocol}

The Bell OMP is a two-step OMP rooted at \(\Omega_0 := \{x\} \subset \mathbb{Z}\). Its complete branch label is $m=(n,j,i,\ell)\in\mathbb{Z}_2^4$, where \(n\) is the final root readout outcome, \(j\) and \(i\) are the first-layer outcomes on \(x-1\) and \(x+1\), respectively, and \(\ell\) is the second-layer outcome on \(x+2\). The readout layer on \(\Omega_0=\{x\}\) is the computational-basis PVM:
\[
    \texttt{J}_{\Omega_0}
    =
    \left\{
        \mathcal J_{\Omega_0}^{(n)}
    \right\}_{n\in\mathbb{Z}_2},
    \qquad
    \mathcal J_{\Omega_0}^{(n)}
    =
    \operatorname{Ad}[\proj{n}_{\{x\}}].
\] The first instrument layer is supported on $\Omega_1=\{x-1,x,x+1\}$. The single-site quantum instruments are
\[
    \texttt{J}_{\{x-1\}_1}
    =
    \Big\{\operatorname{Ad}[\proj{j}_{\{x-1\}}]\Big\}_{j\in\mathbb{Z}_2},
    \quad \mathcal J_{\{x\}_1}
    =
    \operatorname{Ad}[H_{\{x\}}], \quad \texttt{J}_{\{x+1\}_1}
    =
    \Big\{\operatorname{Ad}[\proj{i}_{\{x+1\}}H_{\{x+1\}}]\Big\}_{i\in\mathbb{Z}_2}.
\] The joint first-layer branch is the tensor product
\(
    \mathcal J_{\Omega_1}^{(j,i)}
    :=
    \mathcal J_{\{x-1\}_1}^{(j)}
    \otimes
    \mathcal J_{\{x\}_1}
    \otimes
    \mathcal J_{\{x+1\}_1}^{(i)}
\). The second instrument layer is supported on
\(
    \Omega_2=\{x-2,x-1,x,x+1,x+2\}
\). All sites carry the identity operation except \(x+1\) and \(x+2\). On \(x+1\) we apply the Hadamard channel, and on \(x+2\) we apply the computational-basis PVM:
\[
    \mathcal J_{\{x+1\}_2}
    =
    \operatorname{Ad}[H_{\{x+1\}}],
    \qquad 
    \texttt{J}_{\{x+2\}_2}
    =
    \Big\{\operatorname{Ad}[\proj{\ell}_{\{x+2\}}]\Big\}_{\ell\in\mathbb{Z}_2}.
\] The joint second-layer branch is
\(
    \mathcal J_{\Omega_2}^{(\ell)}
    :=
    \operatorname{id}_{\{x-2\}_2}
    \otimes\operatorname{id}_{\{x-1\}_2}
    \otimes\operatorname{id}_{\{x\}_2}
    \otimes\mathcal J_{\{x+1\}_2}
    \otimes\mathcal J_{\{x+2\}_2}^{(\ell)}.
\) With the Heisenberg ordering used in the main text, the branch map associated with \((n,j,i,\ell)\) is
\[
    \mathcal{P}_{\QCA{q}}^{(n,j,i,\ell)}
    =
    \mathcal J_{\Omega_2}^{(\ell)}
    \circ \QCA{q}
    \circ
    \mathcal J_{\Omega_1}^{(j,i)}
    \circ \QCA{q}
    \circ
    \mathcal J_{\Omega_0}^{(n)} .
\] The effective POVM element is $ E_{\Omega_2}^{(n,j,i,\ell)} := \mathcal{P}_{\QCA{q}}^{(n,j,i,\ell)}(\mathbf 1)$. Therefore, for every initial state \(\omega\) on the quasi-local algebra, the probability of obtaining the outcome $(n,j,i,\ell)$ is 
\[ 
p(n,j,i,\ell|\omega) = \omega(E_{\Omega_2}^{(n,j,i,\ell)}).
\]

At the special value $\varphi=\pi$, the controlled-phase interaction is the controlled-\(Z\) gate, $\operatorname{CPZ}(\pi)=\operatorname{CZ} = \operatorname{diag}(1,1,1,-1)$. Since \(\operatorname{CZ}\) is self-adjoint and diagonal, computational-basis projectors commute with the first backward pullback. Thus
\(
    \QCA{q}(\proj{n}_{\{x\}})
    =
    \proj{n}_{\{x\}}
\). After the first instrument layer one obtains
\[
    \mathcal J_{\Omega_1}^{(j,i)}
    \left(\proj{n}_{\{x\}}\right)
    =
    \proj{j}_{\{x-1\}}
    \otimes
    \proj{\xi_n}_{\{x\}}
    \otimes
    \proj{\xi_i}_{\{x+1\}},
\] where $\ket{\xi_0}:=\ket{+}$ and $\ket{\xi_1}:=\ket{-}$. The second QCA step entangles the two central sites. We first apply the second QCA step. The controlled-\(Z\) gates coupling the pair $\{x,x+1\}$ to the outer sites act conditionally on the outer computational-basis outcomes. Indeed,
\[
    \proj{j}_{\{x-1\}}
    \operatorname{CZ}_{\{x-1,x\}}
    =
    \proj{j}_{\{x-1\}} Z_{\{x\}}^j,
\] while the outcome \(\ell\) registered on \(x+2\) in the
second instrument layer gives
\[
    \operatorname{CZ}_{\{x+1,x+2\}}
    \proj{\ell}_{\{x+2\}}
    =
    Z_{\{x+1\}}^\ell
    \proj{\ell}_{\{x+2\}}.
\] Since \(Z^r\ket{\xi_s}=\ket{\xi_{r\oplus s}}\), these two
conditional operations replace the central input projectors by
\(\proj{\xi_{n\oplus j}}_{\{x\}}\) and \(\proj{\xi_{i\oplus\ell}}_{\{x+1\}}\). The central controlled-\(Z\) gate acts during the QCA step;
subsequently, the second instrument layer applies the Hadamard
on \(x+1\). Thus the resulting central effect is
\[
    H_{\{x+1\}}
    \operatorname{CZ}_{\{x,x+1\}}
    \left(
        \proj{\xi_{n\oplus j}}_{\{x\}}
        \otimes
        \proj{\xi_{i\oplus\ell}}_{\{x+1\}}
    \right)
    \operatorname{CZ}_{\{x,x+1\}}
    H_{\{x+1\}}.
\] To identify this effect, use
\[
    \operatorname{CNOT}_{x\to x+1}
    =
    H_{\{x+1\}}
    \operatorname{CZ}_{\{x,x+1\}}
    H_{\{x+1\}}.
\] Because \(\ket{\xi_b}=H\ket{b}\), we obtain
\[
\begin{aligned}
    H_{\{x+1\}}
    \operatorname{CZ}_{\{x,x+1\}}
    \bigl(\ket{\xi_a}_{\{x\}}\otimes
          \ket{\xi_b}_{\{x+1\}}\bigr) 
          & =
    \operatorname{CNOT}_{x\to x+1}
    \bigl(\ket{\xi_a}_{\{x\}}\otimes
          \ket{b}_{\{x+1\}}\bigr) \\[1mm] 
          & =
    \frac{1}{\sqrt{2}}\Big(
        \ket{0,b}
        +(-1)^a\ket{1,b\oplus1}\Big)
    =:\ket{\beta^{(a,b)}}.
\end{aligned}
\] Consequently,
\[
    \operatorname{Ad}\left[
        \operatorname{CZ}_{\{x,x+1\}}H_{\{x+1\}}
    \right]
    \left(
        \proj{\xi_a}_{\{x\}}
        \otimes
        \proj{\xi_b}_{\{x+1\}}
    \right)
    =
    \proj{\beta^{(a,b)}}_{\{x,x+1\}}.
\] where the Bell basis is
\[
    \ket{\beta^{(0,0)}}
    :=
    \ket{\Phi^+},
    \quad
    \ket{\beta^{(0,1)}}
    :=
    \ket{\Psi^+}, 
    \quad 
    \ket{\beta^{(1,0)}}
    :=
    \ket{\Phi^-},
    \quad
    \ket{\beta^{(1,1)}}
    :=
    \ket{\Psi^-},
\] with
\[
    \ket{\Phi^\pm}
    =
    \frac{\ket{00}\pm\ket{11}}{\sqrt2},
    \qquad
    \ket{\Psi^\pm}
    =
    \frac{\ket{01}\pm\ket{10}}{\sqrt2}.
\] Thus the fine-grained branch effects are
\[
    E_{\Omega_2}^{(n,j,i,\ell)}
    =
    \mathbf{1}_{\{x-2\}} 
    \otimes
    \proj{j}_{\{x-1\}}
    \otimes
    \proj{\beta^{(n\oplus j,i\oplus \ell)}}_{\{x,x+1\}}
    \otimes
    \proj{\ell}_{\{x+2\}}.
\] Define the classical coarse-graining map
\[
    F:\mathbb{Z}_2^4 \to \mathbb{Z}_2^2,
    \qquad
    F(n,j,i,\ell)
    :=
    (n\oplus j,i\oplus \ell).
\] For each \((a,b)\in\mathbb{Z}_2^2\), define the coarse-grained effect
\[
    E_{\Omega_2}^{(a,b)}
    :=
    \sum_{(n,j,i,\ell)\in F^{-1}(a,b)}
    E_{\Omega_2}^{(n,j,i,\ell)} .
\] The condition \(F(n,j,i,\ell)=(a,b)\) is equivalent to
\(n\oplus j=a\) and \(i\oplus\ell=b\). Thus \(j\) and
\(\ell\) can be chosen freely, while \(n=a\oplus j\) and
\(i=b\oplus\ell\) are fixed by these choices. Hence
\[
    F^{-1}(a,b)
    =
    \left\{
        (a\oplus j,\;j,\;b\oplus\ell,\;\ell)
        \mid j,\ell\in\mathbb Z_2
    \right\}.
\] Therefore, after discarding the trivial record sites, the effective POVM on the central pair is exactly the Bell PVM:
\[
    E_{\Omega_2}^{(a,b)}
    =
    \mathbf 1_{\{x-2,x-1\}}
    \otimes
    \proj{\beta^{(a,b)}}_{\{x,x+1\}}
    \otimes
    \mathbf 1_{\{x+2\}}.
    \qquad 
    E_{\Omega_2}^{(a,b)}E_{\Omega_2}^{(a',b')}
    =
    \delta_{a,a'}\delta_{b,b'}E_{\Omega_2}^{(a,b)},
    \qquad 
    \sum_{a,b\in\mathbb{Z}_2}
    E_{\Omega_2}^{(a,b)}
    =
    \mathbf 1_{\Omega_2}.
\] Thus the Bell OMP realizes a projective Bell measurement on two neighbouring sites using only single-site instruments interleaved with causal QCA evolution.

\smallskip
We now consider a general interaction phase
\(\varphi\in(-\pi,\pi]\). The outer outcomes \(j\) and \(\ell\)
still act as controls on the central sites. For a controlled-phase
gate, conditioning one qubit on the computational-basis outcome
\(r\in\{0,1\}\) induces on the other qubit the diagonal operator
\(\operatorname{diag}(1,e^{ir\varphi})\). Up to a global phase,
this is the rotation \(R_Z(\varphi)^r\), where
\(R_Z(\theta):=e^{-i\theta Z/2}\). The two conditional rotations and the controlled-phase gate coupling \(x\) to \(x+1\) therefore give the central unitary
\[
    U^{(\ell,j)}(\varphi)_{\{x,x+1\}}
    :=
    \operatorname{CPZ}(\varphi)_{\{x,x+1\}}
    R_Z(\varphi)_{\{x\}}^j
    R_Z(\varphi)_{\{x+1\}}^\ell .
\] All three factors are diagonal in the computational basis and
hence commute. In the ordered basis
\(\{\ket{00},\ket{01},\ket{10},\ket{11}\}\), their product is
\[
    U^{(\ell,j)}(\varphi)
    =
    e^{-i(j+\ell)\varphi/2}
    \operatorname{diag}\!\left(
        1,\,
        e^{i\ell\varphi},\,
        e^{ij\varphi},\,
        e^{i(1+j+\ell)\varphi}
    \right).
\] The scalar factor does not affect the associated Heisenberg action. 

At \(\varphi=\pi\), the central interaction is \(\operatorname{CPZ}(\pi)=\operatorname{CZ}\), while \(R_Z(\pi)=-iZ\). Consequently,
\[
    U^{(\ell,j)}(\pi)
    =
    (-i)^{j+\ell}
    \operatorname{CZ}\,
    Z_{\{x\}}^j Z_{\{x+1\}}^\ell .
\] This is the Bell-generating unitary, including its global phase. To separate this special value from the deformation, set
\(\delta:=\varphi-\pi\). The identities
\(\operatorname{CPZ}(\theta_1+\theta_2)
 =\operatorname{CPZ}(\theta_1)\operatorname{CPZ}(\theta_2)\)
and
\(R_Z(\theta_1+\theta_2)
 =R_Z(\theta_1)R_Z(\theta_2)\)
give the exact operator identity
\[
    U^{(\ell,j)}(\varphi)
    =
    U^{(\ell,j)}(\delta)U^{(\ell,j)}(\pi).
\] Let \(a:=n\oplus j\) and \(b:=i\oplus\ell\). By the Bell-basis
convention established above,
\[
    \ket{\beta^{(a,b)}}
    =
    H_{\{x+1\}}
    \operatorname{CZ}_{\{x,x+1\}}
    \bigl(
        \ket{\xi_a}_{\{x\}}
        \otimes
        \ket{\xi_b}_{\{x+1\}}
    \bigr).
\] In particular,
\[
    H_{\{x+1\}}\ket{\beta^{(a,b)}}
    =
    \operatorname{CZ}_{\{x,x+1\}}
    \bigl(
        \ket{\xi_a}_{\{x\}}
        \otimes
        \ket{\xi_b}_{\{x+1\}}
    \bigr).
\] We define the two--qubit state associated with the branch
\((n,j,i,\ell)\) by
\[
    \ket{\tau_{\varphi}^{(n\oplus j,i\oplus\ell)}}
    :=
    H_{\{x+1\}}
    U^{(\ell,j)}(\varphi-\pi)_{\{x,x+1\}}
    H_{\{x+1\}}
    \ket{\beta^{(n\oplus j,i\oplus\ell)}}.
\] To see how this state arises from the protocol, apply \(U^{(\ell,j)}(\pi)\) to the states obtained after the first instrument layer. Using \(Z^r\ket{\xi_s}=\ket{\xi_{r\oplus s}}\), we find
\[
\begin{aligned}
    U^{(\ell,j)}(\pi)
    \bigl(
        \ket{\xi_n}_{\{x\}}
        \otimes\ket{\xi_i}_{\{x+1\}}
    \bigr)
    &=
    (-i)^{j+\ell}
    \operatorname{CZ}_{\{x,x+1\}}
    \bigl(
        \ket{\xi_{n\oplus j}}_{\{x\}}
        \otimes
        \ket{\xi_{i\oplus\ell}}_{\{x+1\}}
    \bigr).
\end{aligned}
\]
Combining this equation with the factorization of
\(U^{(\ell,j)}(\varphi)\) yields
\[
    H_{\{x+1\}}U^{(\ell,j)}(\varphi)
    \bigl(
        \ket{\xi_n}_{\{x\}}
        \otimes\ket{\xi_i}_{\{x+1\}}
    \bigr)
    =
    (-i)^{j+\ell}
    \ket{\tau_{\varphi}^{(n\oplus j,i\oplus\ell)}}.
\]
The phase on the right disappears upon taking the projector.
The complete fine-grained effect is therefore
\[
    E_{\Omega_2}^{(n,j,i,\ell)}(\varphi)
    =
    \mathbf 1_{\{x-2\}}
    \otimes
    \proj{j}_{\{x-1\}}
    \otimes
    \proj{\tau_{\varphi}^{(n\oplus j,i\oplus\ell)}}_{\{x,x+1\}}
    \otimes
    \proj{\ell}_{\{x+2\}}.
\]

At \(\varphi=\pi\), we have \(\delta=0\) and
\(U^{(\ell,j)}(0)=\mathbf 1\), then $\ket{\tau_{\pi}^{(n\oplus j,i\oplus\ell)}} = \ket{\beta^{(n\oplus j,i\oplus\ell)}}$, and the Bell PVM is recovered after coarse-graining. For general \(\varphi\), each fixed pair of outer outcomes \((j,\ell)\) specifies a deformed orthonormal basis of central states.

In addition, since \(R_X(\theta):=HR_Z(\theta)H = \exp{(-\mathrm{i}\theta X/2)}\), then 
\[
    \operatorname{CPX}(\theta)_{\{x,x+1\}}
    :=
    H_{\{x+1\}}
    \operatorname{CPZ}(\theta)_{\{x,x+1\}}
    H_{\{x+1\}}.
\] Conjugating the three commuting factors of
\(U^{(\ell,j)}(\theta)\) by \(H_{\{x+1\}}\) gives
\[
    H_{\{x+1\}}U^{(\ell,j)}(\theta)H_{\{x+1\}}
    =
    \operatorname{CPX}(\theta)_{\{x,x+1\}}
    R_Z(\theta)_{\{x\}}^j
    R_X(\theta)_{\{x+1\}}^\ell .
\] Thus, with \(\theta=\varphi-\pi\), 
\[
    \ket{\tau_{\varphi}^{(n\oplus j,i\oplus\ell)}}
    =
    \operatorname{CPX}(\theta)_{\{x,x+1\}}
    R_Z(\theta)_{\{x\}}^j
    R_X(\theta)_{\{x+1\}}^\ell
    \ket{\beta^{(n\oplus j,i\oplus\ell)}}.
\]

\subsection{Entanglement Entropy of the OMP Branches}

We first determine the entanglement of the central state associated
with a fine-grained outcome \((n,j,i,\ell)\). Throughout this
subsection, \(n,j,i,\ell\in\mathbb Z_2\). As established above, the
central projector in the corresponding effect is a projection onto \(\ket{\tau_\varphi^{(n\oplus j,i\oplus\ell)}}\), with the dependence on \(j\) and \(\ell\) also entering through \(U^{(\ell,j)}(\varphi-\pi)\).

It is convenient to calculate the entanglement using the state
\[
    \ket{\Psi_\varphi^{(j,n,\ell,i)}}
    :=
    \operatorname{CPZ}(\varphi)
    \left(
        R_Z(\varphi)^j H\ket{n}
        \otimes
        R_Z(\varphi)^\ell H\ket{i}
    \right).
\] The relation between this state and $\ket{\tau_\varphi^{(n\oplus j,i\oplus\ell)}}$. Put \(a=n\oplus j\), \(b=i\oplus\ell\), and
\(\delta=\varphi-\pi\). Since
\[
    \ket{\beta^{(a,b)}}
    =
    (\mathbf 1\otimes H)\operatorname{CZ}
    \bigl(\ket{\xi_a}\otimes\ket{\xi_b}\bigr),
\] the definition of the deformed Bell state gives
\[
\begin{aligned}
    \ket{\tau_\varphi^{(n\oplus j,i\oplus\ell)}}
    &=
    (\mathbf 1\otimes H)
    U^{(\ell,j)}(\delta)\operatorname{CZ}
    \bigl(
        \ket{\xi_{n\oplus j}}
        \otimes
        \ket{\xi_{i\oplus\ell}}
    \bigr)\\                                                   
    &=
    (\mathbf 1\otimes H)
    U^{(\ell,j)}(\delta)\operatorname{CZ}
    (Z^j\otimes Z^\ell)
    \bigl(\ket{\xi_n}\otimes\ket{\xi_i}\bigr) \\
    &=
    (\mathbf 1\otimes H)
    U^{(\ell,j)}(\delta)\operatorname{CZ}
    (Z^j\otimes Z^\ell)
    \bigl(H\ket{n} \otimes H\ket{i}\bigr).
\end{aligned}
\] Here we used \(Z^r\ket{\xi_s}=\ket{\xi_{r\oplus s}}\). Furthermore, \(R_Z(\pi)=-iZ\) and \(U^{(\ell,j)}(\varphi)
 =U^{(\ell,j)}(\delta)U^{(\ell,j)}(\pi)\), so that
\[
    U^{(\ell,j)}(\delta)\operatorname{CZ}
    (Z^j\otimes Z^\ell)
    =
    i^{j+\ell}U^{(\ell,j)}(\varphi).
\] Consequently,
\[
\begin{aligned}
    \ket{\tau_\varphi^{(n\oplus j,i\oplus\ell)}}
    & = i^{j+\ell}(\mathbf 1\otimes H)
    U^{(\ell,j)}(\varphi)\bigl(H\ket{n} \otimes H\ket{i}\bigr) \\
    & = i^{j+\ell}(\mathbf 1\otimes H)
    \operatorname{CPZ}(\varphi)
    \bigr(R_Z(\varphi)^j \otimes
    R_Z(\varphi)^\ell\bigl)\bigl(H\ket{n} \otimes H\ket{i}\bigr) \\
    & = i^{j+\ell}
    (\mathbf 1\otimes H)
    \ket{\Psi_\varphi^{(j,n,\ell,i)}}.
\end{aligned}
\] The phase \(i^{j+\ell}\) has no effect on the projector, and the
Hadamard acts locally on the second qubit. Hence the two states
have the same entanglement entropy. Write
\[
    u:=j\varphi+n\pi,
    \qquad
    v:=\ell\varphi+i\pi.
\]
Since
\[
    R_Z(\varphi)^jH\ket{n}
    =
    \frac{e^{-ij\varphi/2}}{\sqrt2}
    \left(\ket{0}+e^{iu}\ket{1}\right),
    \qquad
    R_Z(\varphi)^\ell H\ket{i}
    =
    \frac{e^{-i\ell\varphi/2}}{\sqrt2}
    \left(\ket{0}+e^{iv}\ket{1}\right),
\] the two inputs carry the combined global phase
\(e^{-i(j+\ell)\varphi/2}\). Applying
\(\operatorname{CPZ}(\varphi)
=\operatorname{diag}(1,1,1,e^{i\varphi})\) gives
\[
\begin{aligned}
    \ket{\Psi_\varphi^{(j,n,\ell,i)}}
    &=
    \frac{e^{-i(j+\ell)\varphi/2}}{2}
    \Bigl(
        \ket{00}
        +e^{iv}\ket{01}
        +e^{iu}\ket{10}
        +e^{i(u+v+\varphi)}\ket{11}
    \Bigr)\\
    &=
    \frac{e^{-i(j+\ell)\varphi/2}}{\sqrt2}
    \left(
        \ket{0}\ket{\chi_0}
        +e^{iu}\ket{1}\ket{\chi_1}
    \right),
\end{aligned}
\]
where
\[
    \ket{\chi_0}
    :=
    \frac{\ket{0}+e^{iv}\ket{1}}{\sqrt2},
    \qquad
    \ket{\chi_1}
    :=
    \frac{\ket{0}+e^{i(v+\varphi)}\ket{1}}{\sqrt2}.
\] Both conditional vectors are normalized. Their overlap is
\[
    \langle\chi_0|\chi_1\rangle
    =
    \frac{1+e^{i\varphi}}{2}
    =
    e^{i\varphi/2}\cos\frac{\varphi}{2}.
\] Thus
\(\bigl|\langle\chi_0|\chi_1\rangle\bigr|
=|\cos(\varphi/2)|\), independently of all four branch
labels. The common factor \(e^{-i(j+\ell)\varphi/2}\) does not enter the reduced density matrix.

Let
\(\rho_\Psi
 :=\proj{\Psi_\varphi^{(j,n,\ell,i)}}\).
Tracing out the first qubit removes the cross terms between
\(\ket{0}\) and \(\ket{1}\), giving
\[
    {\rho_{\Psi}}_{\{x+1\}}
    :=
    \operatorname{Tr}_{\{x\}}[\rho_\Psi]
    =
    \frac12\bigl(
        \proj{\chi_0}+\proj{\chi_1}
    \bigr).
\] Its two eigenvalues are
\[
    \lambda_\pm(\varphi)
    =
    \frac{1\pm|\langle\chi_0|\chi_1\rangle|}{2}
    =
    \frac{1\pm|\cos(\varphi/2)|}{2}.
\]
Because \(\ket{\Psi_\varphi^{(j,n,\ell,i)}}\) is pure,
these are also the squared Schmidt coefficients of the
central PVM state. Its entanglement entropy is therefore
\[
    S(\varphi)
    =
    -\sum_{\sigma\in\{+,-\}}
      \lambda_\sigma(\varphi)
      \log_2\lambda_\sigma(\varphi)
    =
    H_2\left(
        \frac{1+|\cos(\varphi/2)|}{2}
    \right),
\] where $H_2(p):=-p\log_2p-(1-p)\log_2(1-p)$, is the binary entropy, defined with the usual convention \(0\log_2 0=0\). This result is independent of \((n,j,i,\ell)\). In particular, \(S(0)=0\) and \(S(\pi)=1\).

\smallskip
We now compare the Bell OMP inputs with arbitrary product inputs
to \(\operatorname{CPZ}(\varphi)\). Fix the interaction phase
\(\varphi\in(-\pi,\pi]\), and write two normalized one-qubit
states, up to individual global phases, as
\[
    \ket{\psi_k}
    =
    \cos\frac{\vartheta_k}{2}\ket{0}
    +
    e^{i\xi_k}\sin\frac{\vartheta_k}{2}\ket{1},
    \qquad
    0\leq\vartheta_k\leq\pi,
    \quad k=1,2.
\] Set
\[
    \ket{\Psi}
    :=
    \operatorname{CPZ}(\varphi)
    \bigl(\ket{\psi_1}\otimes\ket{\psi_2}\bigr)
    =
    \sum_{a,b\in\mathbb Z_2}C_{ab}\ket{ab}.
\] The gate multiplies only the coefficient of \(\ket{11}\)
by \(e^{i\varphi}\). Therefore
\[
\begin{aligned}
    C_{00}
    &=
    \cos\frac{\vartheta_1}{2}
    \cos\frac{\vartheta_2}{2},
    &
    C_{01}
    &=
    e^{i\xi_2}
    \cos\frac{\vartheta_1}{2}
    \sin\frac{\vartheta_2}{2},
    \\
    C_{10}
    &=
    e^{i\xi_1}
    \sin\frac{\vartheta_1}{2}
    \cos\frac{\vartheta_2}{2},
    &
    C_{11}
    &=
    e^{i(\xi_1+\xi_2+\varphi)}
    \sin\frac{\vartheta_1}{2}
    \sin\frac{\vartheta_2}{2}.
\end{aligned}
\] Regard the amplitudes as the entries of the matrix
\[
    C=
    \begin{pmatrix}
        C_{00}&C_{01}\\
        C_{10}&C_{11}
    \end{pmatrix}.
\] Tracing out the first qubit gives
\(\rho_{\{x+1\}}=C^\dagger C\). Hence
\[
    \det\rho_{\{x+1\}}
    =
    \det(C^\dagger C)
    =
    |\det C|^2
    =
    |C_{00}C_{11}-C_{01}C_{10}|^2.
\] Substitution of the four amplitudes yields
\[
\begin{aligned}
    C_{00}C_{11}-C_{01}C_{10}
    &=
    e^{i(\xi_1+\xi_2)}
    \cos\frac{\vartheta_1}{2}
    \sin\frac{\vartheta_1}{2}
    \cos\frac{\vartheta_2}{2}
    \sin\frac{\vartheta_2}{2}
    \bigl(e^{i\varphi}-1\bigr)\\
    &=
    \frac14 e^{i(\xi_1+\xi_2)}
    \sin\vartheta_1\sin\vartheta_2
    \bigl(e^{i\varphi}-1\bigr).
\end{aligned}
\] Since
\(|e^{i\varphi}-1|^2=4\sin^2(\varphi/2)\),
we obtain
\[
    \det\rho_{\{x+1\}}
    =
    \frac14
    \sin^2\vartheta_1
    \sin^2\vartheta_2
    \sin^2\frac{\varphi}{2}.
\] The phases \(\xi_1,\xi_2\) have disappeared because they
contribute only the unit-modulus factor
\(e^{i(\xi_1+\xi_2)}\) to \(\det C\). The reduced state is a \(2\times2\) density matrix of unit trace. Its eigenvalues therefore solve \(t^2-t+\det\rho_{\{x+1\}}=0\), and are
\[
    \lambda_\pm
    =
    \frac12\left(
        1\pm
        \sqrt{
            1-
            \sin^2\vartheta_1
            \sin^2\vartheta_2
            \sin^2\frac{\varphi}{2}
        }
    \right).
\]
Put
\[
    s:=
    \sqrt{
        1-
        \sin^2\vartheta_1
        \sin^2\vartheta_2
        \sin^2\frac{\varphi}{2}
    }.
\]
The entanglement entropy of \(\ket{\Psi}\) is
\(H_2((1+s)/2)\). For \(0<s<1\),
\[
    \frac{\mathrm d}{\mathrm d s}
    H_2\left(\frac{1+s}{2}\right)
    =
    \frac12\log_2\frac{1-s}{1+s}
    <0.
\] By continuity the same monotonicity holds at the endpoints. Thus, at fixed \(\varphi\), maximizing the entanglement is equivalent to minimizing \(s\), or equivalently to maximizing \(\sin^2\vartheta_1\sin^2\vartheta_2\). Because \(0\leq\sin^2\vartheta_k\leq1\), the largest possible value of this product is \(1\), attained at \(\vartheta_1=\vartheta_2=\pi/2\). These are precisely equatorial inputs. When
\(\sin(\varphi/2)\neq0\), this condition is necessary
as well as sufficient for maximal entanglement.
When \(\sin(\varphi/2)=0\), the gate is the identity
within the phase range considered, so every product
input produces zero entanglement and every choice of
\(\vartheta_1,\vartheta_2\) is a maximizer. Finally, the Bell OMP inputs
\(R_Z(\varphi)^jH\ket{n}\) and
\(R_Z(\varphi)^\ell H\ket{i}\) both have computational-basis
probabilities \((1/2,1/2)\). They therefore satisfy
\(\vartheta_1=\vartheta_2=\pi/2\), independently of the
branch labels. Every branch attains the maximal
entanglement obtainable from product inputs at its
fixed interaction phase.

\subsection{Stabilizer Rényi Entropy of the OMP Branches}

We next evaluate the stabilizer Rényi entropy \cite{LeoneOlivieroHamma2022} of the states
onto which the protocol effects project. Here the relation
between \(\ket{\tau_\varphi^{(n\oplus j,i\oplus\ell)}}\)
and the representative \(\ket{\Psi_\varphi^{(j,n,\ell,i)}}\) must be checked at the level of Clifford equivalence. The identity established in the preceding subsection is
\[
    \ket{\tau_\varphi^{(n\oplus j,i\oplus\ell)}}
    =
    i^{j+\ell}
    (\mathbf 1\otimes H)
    \ket{\Psi_\varphi^{(j,n,\ell,i)}}.
\] Since \(H\) is a Clifford unitary, the two states have the
same stabilizer Rényi entropy. Indeed, Clifford conjugation
permutes Pauli operators modulo phases. If \(C\) is Clifford,
then \(C\mathbb{P}_NC^\dagger=\mathbb{P}_N\), and the
sum defining the entropy is unchanged under
\(\ket{\psi}\mapsto C\ket{\psi}\). Thus the calculation on
\(\ket{\Psi_\varphi^{(j,n,\ell,i)}}\) gives exactly the
entropy of the central PVM state.

Let \(\mathbb{P}_N=\{I,X,Y,Z\}^{\otimes N}\) denote the
\(N\)-qubit Pauli group modulo phases. For a pure state
\(\ket{\psi}\), define
\[
    \Xi_P[\ket{\psi}]
    :=
    \frac{
        |\langle\psi|P|\psi\rangle|^2
    }{2^N},
    \qquad P\in\mathbb{P}_N.
\]
The stabilizer Rényi entropy of order
\(\alpha>0\), \(\alpha\neq1\), is
\[
    M_\alpha[\ket{\psi}]
    :=
    \frac{1}{1-\alpha}
    \log_2\!\left(
        \sum_{P\in\mathbb{P}_N}
        \Xi_P[\ket{\psi}]^\alpha
    \right)-N.
\] For two qubits and \(\alpha=2\), this reduces to
\[
    M_2[\ket{\psi}]
    =
    2-\log_2\!\left(
        \sum_{P\in\mathbb{P}_2}
        |\langle\psi|P|\psi\rangle|^4
    \right).
\] We calculate the sixteen Pauli expectation values from the
explicit amplitudes of the representative. Set
\[
    u:=j\varphi+n\pi,
    \qquad
    v:=\ell\varphi+i\pi.
\]
Removing an irrelevant global phase, its amplitudes in the
computational basis are
\[
    C_{00}=\frac12,\qquad
    C_{01}=\frac{e^{iv}}2,\qquad
    C_{10}=\frac{e^{iu}}2,\qquad
    C_{11}=\frac{e^{i(u+v+\varphi)}}2.
\]
For \(P,Q\in\{I,X,Y,Z\}\), write
\[
    \langle P\otimes Q\rangle_\Psi
    :=
    \langle\Psi_\varphi^{(j,n,\ell,i)}
      |P\otimes Q|
      \Psi_\varphi^{(j,n,\ell,i)}\rangle.
\]
The matrix elements give the general expansion
\[
    \langle P\otimes Q\rangle_\Psi
    =
    \sum_{a,b,c,d\in\mathbb Z_2}
    \overline{C_{ab}}\,C_{cd}\,
    \langle a|P|c\rangle
    \langle b|Q|d\rangle.
\]
For the amplitudes above, the diagonal expectations are
\[
    \langle I\otimes I\rangle_\Psi=1,
    \qquad
    \langle Z\otimes I\rangle_\Psi
    =
    \langle I\otimes Z\rangle_\Psi
    =
    \langle Z\otimes Z\rangle_\Psi=0.
\]
The expectations with one \(X\) or \(Y\) and one identity
are
\[
\begin{aligned}
    \langle X\otimes I\rangle_\Psi
    &=
    \frac{\cos u+\cos(u+\varphi)}2,
    &
    \langle Y\otimes I\rangle_\Psi
    &=
    \frac{\sin u+\sin(u+\varphi)}2,\\
    \langle I\otimes X\rangle_\Psi
    &=
    \frac{\cos v+\cos(v+\varphi)}2,
    &
    \langle I\otimes Y\rangle_\Psi
    &=
    \frac{\sin v+\sin(v+\varphi)}2.
\end{aligned}
\]
Replacing the identity by \(Z\) changes the relative sign:
\[
\begin{aligned}
    \langle X\otimes Z\rangle_\Psi
    &=
    \frac{\cos u-\cos(u+\varphi)}2,
    &
    \langle Y\otimes Z\rangle_\Psi
    &=
    \frac{\sin u-\sin(u+\varphi)}2,\\
    \langle Z\otimes X\rangle_\Psi
    &=
    \frac{\cos v-\cos(v+\varphi)}2,
    &
    \langle Z\otimes Y\rangle_\Psi
    &=
    \frac{\sin v-\sin(v+\varphi)}2.
\end{aligned}
\]
Finally, the four expectations involving two transverse
Pauli operators are
\[
\begin{aligned}
    \langle X\otimes X\rangle_\Psi
    &=
    \frac{
        \cos(u+v+\varphi)+\cos(u-v)
    }2,
    &
    \langle X\otimes Y\rangle_\Psi
    &=
    \frac{
        \sin(u+v+\varphi)-\sin(u-v)
    }2,\\
    \langle Y\otimes X\rangle_\Psi
    &=
    \frac{
        \sin(u+v+\varphi)+\sin(u-v)
    }2,
    &
    \langle Y\otimes Y\rangle_\Psi
    &=
    \frac{
        \cos(u-v)-\cos(u+v+\varphi)
    }2.
\end{aligned}
\]
These expressions give all sixteen terms entering \(M_2\). 

\smallskip

Before summing their fourth powers, the dependence on the readout labels can be removed exactly. Changing \(n\) or \(i\)
acts on the corresponding local input by \(Z\):
\[
    R_Z(\varphi)^jH\ket{1}
    =
    ZR_Z(\varphi)^jH\ket{0},
    \qquad
    R_Z(\varphi)^\ell H\ket{1}
    =
    ZR_Z(\varphi)^\ell H\ket{0}.
\] Both local \(Z\) operators commute with
\(\operatorname{CPZ}(\varphi)\). Hence changing \(n\) or \(i\)
applies a local Pauli, and therefore a Clifford, to the
two-qubit representative. Its stabilizer Rényi entropy
depends only on \((j,\ell)\). Define
\[
    \mathcal S^{(j,\ell)}(\varphi)
    :=
    \sum_{P\in\mathbb{P}_2}
    \left|
        \langle
        \Psi_\varphi^{(j,n,\ell,i)}
        |P|
        \Psi_\varphi^{(j,n,\ell,i)}
        \rangle
    \right|^4.
\]
The preceding argument shows that the right-hand side is
independent of \(n\) and \(i\). Substituting
\(u=j\varphi+n\pi\) and \(v=\ell\varphi+i\pi\) into the
sixteen expectations and putting \(c:=\cos\varphi\), one
obtains
\[
    \mathcal S^{(0,0)}(\varphi)
    =
    1+\frac34(1+c^2)^2, \qquad \mathcal S^{(0,1)}(\varphi)
    =
    \mathcal S^{(1,0)}(\varphi)
    =
    \frac14
    \left(
        7+22c^2-29c^4-16c^6+32c^8
    \right),
\]
and
\[
\begin{aligned}
    \mathcal S^{(1,1)}(\varphi)
    =
    \frac74+\frac{19}{2}c^2+\frac34c^4
    -104c^6+224c^8 -192c^{10}+64c^{12}.
\end{aligned}
\] Therefore the entropy of the central PVM state in each
branch is
\[
    M_2^{(j,\ell)}(\varphi)
    =
    2-\log_2\mathcal S^{(j,\ell)}(\varphi).
\] At \(\varphi=0\), all local inputs are stabilizer states
and \(\operatorname{CPZ}(0)=\mathbf 1\), giving
\(M_2^{(j,\ell)}(0)=0\). At the Bell point,
\(\operatorname{CPZ}(\pi)=\operatorname{CZ}\) is Clifford,
and \(M_2^{(j,\ell)}(\pi)=0\). At
\(\varphi=\pm\pi/2\), all three expressions reduce to
\(\mathcal S^{(j,\ell)}=7/4\); hence
\[
    M_2^{(j,\ell)}
    \!\left(\pm\frac{\pi}{2}\right)
    =
    2-\log_2\frac74
    =
    \log_2\frac{16}{7}.
\]
For \(c^2\in[0,1]\), the three polynomials above satisfy
\(\mathcal S^{(j,\ell)}(\varphi)\geq 7/4\).
Thus, within each branch class, this value is the maximum
attained along the family:
\[
    \max_{\varphi\in(-\pi,\pi]}
    M_2^{(j,\ell)}(\varphi)
    =
    \log_2\frac{16}{7}.
\]

\end{document}